\documentclass[fleqn,usenatbib,useAMS]{Papers}
\usepackage{newtxtext,newtxmath}
\usepackage[T1]{fontenc}
\usepackage{subfloat}
\usepackage{makecell}
\usepackage[authoryear]{natbib}
\usepackage{float}
\usepackage{graphicx}	
\usepackage{amsmath}	

\title[Inverse energy cascade in dark matter flow]{Inverse energy cascade in self-gravitating collisionless dark matter flow and effects of halo shape}

\author[Z. Xu]{
Zhijie (Jay) Xu,$^{1}$\thanks{E-mail: zhijie.xu@pnnl.gov; zhijiexu@hotmail.com}
\\
$^{1}$Physical and Computational Sciences Directorate, Pacific Northwest National Laboratory; Richland, WA 99352, USA\\
}

\date{Accepted XXX. Received YYY; in original form ZZZ}

\pubyear{2022}

\begin{document}
\label{firstpage}
\pagerange{\pageref{firstpage}--\pageref{lastpage}}
\maketitle

\begin{abstract}
Halo-mediated mass and energy cascades are key to understand dark matter flow. Both cascades origin from the mass exchange between halo and out-of-halo sub-systems. Kinetic energy can be from the motion of halos and particle motion in halos. Similarly, potential energy can be due to the inter- and intra-halo interactions. Intra-halo virial equilibrium is established much faster than inter-halo. Change of energy of entire system comes from virilization in halos. At statistically steady state, continuous mass exchange is required to sustain growth of total halo mass $M_h\propto a^{1/2}$ and energy $E\propto a^{3/2}$, where $a$ is scale factor. Inverse cascade is identified for kinetic energy that is transferred from the smallest scale to larger mass scales. This is sustained by the direct cascade of potential energy from large to small scale. Both energies have a scale- and time-independent flux in propagation range that is proportional to mass flux. Energy cascade is mostly facilitated by the mass cascade, which can be quantitatively described by the mass accretion of typical halos. Halo radial, angular momentum, and angular velocity are also modelled and an inverse cascade is identified for the coherent radial and rotational motion in halos. In hydrodynamic turbulence, vortex stretching (shape changing) along its axis of spin enables energy cascade from large to small length scales. However, change in halo shape is not the dominant mechanism for energy cascade as the halo moment of inertial gained from shape changing is less than 2 times. Large halos exhibit preference for prolateness over oblateness and most halos have spin axis perpendicular to major axis. Since mass cascade is local in mass space, halo shape evolves continuously in mass space with large halos formed by incrementally inheriting structure from their progenitor halos. A unique evolution path of halo structure is found that gradually approaches sphere with increasing size.
\end{abstract}

\begin{keywords}
\vspace*{-10pt}
Dark matter; N-body simulations; Theoretical models
\end{keywords}

\begingroup
\let\clearpage\relax
\tableofcontents
\endgroup

\section{ Introduction}
\label{sec:1}
The self-gravitating collisionless fluid dynamics (SG-CFD) concerns the motion of collisionless matter due to its own gravity. "Collisionless" refers to a matter with extremely low interaction cross-section such that the collisions between particles have no effects on the system dynamics. Typical examples can be the cold dark matter and low-density plasmas under high magnetic field. The gravitational collapse of collisionless dark matter is a good example of the nonlinear SG-CFD problem, where gravitational instability leads to the structure formation and evolution. The inverse mass cascade is a key feature of self-gravitating collisionless dark matter flow (SG-CFD) that is not present in the incompressible hydrodynamic turbulence \citep{Xu:2021-Inverse-mass-cascade-mass-function,Xu:2021-Inverse-mass-cascade-halo-density}. However, both dark matter flow and turbulence involve energy cascade across different scales.

Turbulence is ubiquitous in nature that represents one of the most challenging and fascinating problems in classical physics, where the difficulty mostly stems from the random and nonlinear nature with inherent presence of many inseparable scales. To understand the energy cascade in SG-CFD, it is beneficial by revisiting some fundamental ideas in turbulence. The classical picture of turbulence is a cascade process, where large eddies feed smaller eddies, which feed even smaller eddies, and so on to the smallest scale where viscous dissipation is dominant, i.e. the concept of a direct (kinetic) energy cascade \citep{Richardson:1922-Weather-Prediction-by-Numerica}. For incompressible turbulence, the energy cascade starts with the kinetic energy obtained from mean flow by the largest eddies (integral scale) through Reynolds stress (arising from velocity fluctuations). The kinetic energy continuously transferred from mean flow to random motion of turbulence and further transferred successively down to smaller and smaller eddies until the viscous dissipation becomes dominant due to the collisions between molecules. A quantitative description of energy cascade in turbulence based on similarity principles was laid out in seminal works in 1940s \citep{Kolmogoroff:1941-The-local-structure-of-turbule,Kolmogorov:1991-Dissipation-of-Energy-in-the-L}. 

For three-dimensional incompressible flow with high Reynolds number (or negligible viscosity), the energy transfer from mean flow and cascade across scales proceed through a "vortex stretching" mechanism \citep{Taylor:1932-The-transport-of-vorticity-and,Taylor:1938-Production-and-dissipation-of-}. In the presence of shear stress in mean flow, the Reynolds stress due to velocity fluctuation \citep{Andersson:2012-Computational-Fluid-Dynamics-f} acts as a conduit to continuously draw energy from mean flow and sustain the energy cascade in turbulence. The equation for vorticity $\boldsymbol{\mathrm{\omega }}$ simply reads, 
\begin{equation} 
\label{ZEqnNum426779} 
\frac{\partial \boldsymbol{\mathrm{\omega }}}{\partial t} +\left(\boldsymbol{\mathrm{u}}\cdot \nabla \right)\boldsymbol{\mathrm{\omega }}=\underbrace{\left(\boldsymbol{\mathrm{\omega }}\cdot \nabla \right)\boldsymbol{\mathrm{u}}}_{1}+\nu \nabla ^{2} \boldsymbol{\mathrm{\omega }},        
\end{equation} 
where term 1 leads to the vortex stretching. Here $\boldsymbol{\mathrm{u}}$ is the velocity field and vortices' volume is assumed to be conserved for incompressible flow. The shear stress induced lengthening of vortices along the direction of vorticity vector ($\boldsymbol{\mathrm{\omega }}=\nabla \times \boldsymbol{\mathrm{u}}$) implies a thinning of the fluid element in the direction perpendicular to the stretching. This will intensify the vorticity with rising kinetic energy due to the conservation of angular momentum if viscous effect is negligible. 

Let's assume a volume conserved fluid element with moment of inertial $I_{1} $, $I_{2} $ and vorticity $\omega _{1}^{} $, $\omega _{2}^{} $ right before and after vortex stretching. The ratio of kinetic energy from the conservation of angular momentum $I_{1} \omega _{1}^{} =I_{2} \omega _{2}^{} $ is
\begin{equation} 
\label{eq:2} 
\frac{I_{2} \omega _{2}^{2} }{I_{1} \omega _{1}^{2} } =\frac{I_{1} }{I_{2} } ,            
\end{equation} 
where a smaller moment of inertial about the axis of rotating after stretching ($I_{2} <I_{1} $) leads to a greater vorticity ($\omega _{2} >\omega _{1} $) and a rising rotational kinetic energy. With vortices teased out into thinner and thinner filaments, kinetic energy is passed down to smaller and smaller scales and finally dissipated by the molecular viscosity. For high-Re number flow with a vanishing viscosity $\nu \to 0$, enstrophy ($\left|\boldsymbol{\mathrm{\omega }}\right|^{2} $, the square of vorticity) can be unbounded and approach infinity. However, the rate of energy dissipation (proportional to $\nu \left|\boldsymbol{\mathrm{\omega }}\right|^{2} $) is still finite even with $\nu \to 0$ such that the total kinetic energy decreases with time.  

While direct energy cascade is a dominant feature for 3D turbulence, two-dimensional turbulence exhibits an inverse energy cascade (resembles SG-CFD). In fact, there exists a range of length scales over which kinetic energy is transferred from small to large scales for two-dimensional turbulence, i.e. an inverse energy cascade predicted in the late 1960s \citep{Kraichnan:1967-Inertial-Ranges-in-2-Dimension}. The two-dimensional turbulence is not just simply a reduced-dimension version of three-dimensional turbulence. New conservation laws in 2D flow leads to a completely different phenomenology. Note that the vorticity vector $\boldsymbol{\mathrm{\omega }}$ reduces to a scalar $\omega $ in 2D (or always perpendicular to velocity field $\boldsymbol{\mathrm{u}}$), vortex stretching cannot work (term 1 in Eq. \eqref{ZEqnNum426779} is not present in 2D). 

The total enstrophy ($\omega^{2}$) in a 2D flow is destroyed by viscosity and should monotonically decrease with time. Therefore, enstrophy is bounded from above by its initial value. For high-Re two-dimensional incompressible flow with viscosity $\nu \to 0$, the energy dissipation rate $\nu \omega ^{2} \to 0$ which is finite in 3D flow. The enstrophy $\omega ^{2}$ in 2D flow is a bounded quantity such that the total system energy is nearly conserved with energy dissipation rate $\nu \omega ^{2} \to 0$. However, the rate of enstrophy dissipation (proportional to $\nu \left(\nabla \omega \right)^{2} $) can be finite with palinstrophy ($\left(\nabla \omega \right)^{2} $) unbounded and approaching infinity. The kinetic energy ($\left|\boldsymbol{\mathrm{u}}\right|^{2} $) is nearly conserved, while the enstrophy $\omega ^{2} $ is dissipated in high-Re 2D turbulence. The vorticity $\omega$ simply replaces the role of $\boldsymbol{\mathrm{u}}$ in 3D turbulence.  

It turns out that a fully developed 2D turbulence has: a direct cascade of enstrophy ($\omega ^{2} $) from large to small scales and an inverse cascade of kinetic energy ($u^{2}$) from small to large scales. As flow develops, the enstrophy is continually passed down to small scales, where it is destroyed by viscosity. While it is not possible for vortex stretching to operate in two-dimensional flow, vorticity is materially conserved and simply advected like a passive scalar to create a continuous filamentation of vorticity for high-Re incompressible flow. Continuously area-conserved teasing and twisting make vortex patches thinner and longer that facilitates a combined direct cascade of enstrophy and inverse cascade of kinetic energy. 

Based on this brief description of energy/enstrophy cascade in turbulence, let us turn to some unique features of SG-CFD and their fundamental effects on the energy transfer, cascade, and destruction. Specifically, SG-CFD involves a) long-range interaction \citep{Padmanabhan:1990-Statistical-Mechanics-of-Gravi}; b) absence of incompressibility; and c) collisionless nature. The long-range interaction requires the formation of halos (the counterpart of vortex) to maximize system entropy \citep{Xu:2021-The-maximum-entropy-distributi,Xu:2021-Mass-functions-of-dark-matter-}, a unique and elementary structure of SG-CFD \citep{Neyman:1952-A-Theory-of-the-Spatial-Distri,Cooray:2002-Halo-models-of-large-scale-str}. The lack of incompressibility implies a non-zero velocity divergence. On large scale and in the linear regime, the density fluctuations are proportional to velocity divergence that leads to a nonuniform density field. In addition, We can demonstrate that SG-CFD (dark matter flow) is of constant divergence on small scale and is irrotational on large scale \citep{Xu:2022-The-statistical-theory-of-2nd,Xu:2022-The-statistical-theory-of-3rd,Xu:2022-Two-thirds-law-for-pairwise-ve}. Finally, while viscous dissipation is the only mechanism to dissipate kinetic energy and/or enstrophy, it is not present in collisionless SG-CFD. Energy can only be cascaded and transferred in different forms, but not destroyed. 

While inverse mass cascade is not present in hydrodynamic turbulence, it is a key feature of dark matter flow \citep{Xu:2021-Inverse-mass-cascade-mass-function} and is highly correlated with energy cascade (see Eqs. \eqref{ZEqnNum490431} and \eqref{ZEqnNum258650}). There exists a broad spectrum of halo sizes. Halos pass their mass onto larger and larger halos, until mass growth becomes dominant over mass propagation. Consequently, we expect a continuous multistage cascade of mass from smaller to larger mass scales, i.e. "Little halos have big halos, That feed on their mass; And big halos have greater halos, And so on to growth". 

Effects of mass cascade on halo mass function, energy, size and density profile have been previously studied with new mass function proposed \citep{Xu:2021-Inverse-mass-cascade-mass-function,Xu:2021-Inverse-mass-cascade-halo-density}. This paper focus on the energy cascade and its relation to mass cascade. Key questions to address are: i) what is the major mechanism responsible for energy cascade in dark matter flow? ii) Does halo shape play the similar role as "vortex stretching" in turbulence? Answer to these questions might be relevant for dark matter particle mass and properties \citep{Xu:2022-Postulating-dark-matter-partic}, MOND (modified Newtonian dynamics) theory \citep{Xu:2022-The-origin-of-MOND-acceleratio}, and baryonic-to-halo mass relation \citep{Xu:2022-The-baryonic-to-halo-mass-rela}. 

The rest of paper is organized as follows: Section \ref{sec:2} introduces the simulation used for this work. Section \ref{sec:3} briefly reviews the key formulations of inverse mass cascade, followed by developing energy cascade into mathematical formulations in Section \ref{sec:4}. The temporal evolution of total kinetic and potential energies in the entire system is also studied. Effects of halo shape change on energy cascade are discussed in Section \ref{sec:5}. 

\section{N-body simulations and numerical data}
\label{sec:2}
The numerical data for this work is publicly available and generated from the \textit{N}-body simulations carried out by the Virgo consortium. A comprehensive description of the simulation data can be found in \citep{Frenk:2000-Public-Release-of-N-body-simul,Jenkins:1998-Evolution-of-structure-in-cold}. The same set of simulation data has been widely used in a number of different studies from clustering statistics \citep{Jenkins:1998-Evolution-of-structure-in-cold} to the formation of halo clusters in large scale environments \citep{Colberg:1999-Linking-cluster-formation-to-l}, and testing models for halo abundance and mass functions \citep{Sheth:2001-Ellipsoidal-collapse-and-an-im}. More details on simulation parameters are provided in Table \ref{tab:1}.

Two relevant datasets from this N-boby simulation, i.e. a halo-based and correlation-based statistics of dark matter flow, can be found at Zenodo.org  \citep{Xu:2022-Dark_matter-flow-dataset-part1, Xu:2022-Dark_matter-flow-dataset-part2}, along with the accompanying presentation slides "A comparative study of dark matter flow \& hydrodynamic turbulence and its applications" \citep{Xu:2022-Dark_matter-flow-and-hydrodynamic-turbulence-presentation}. All data files are also available on GitHub \citep{Xu:Dark_matter_flow_dataset_2022_all_files}. 

\begin{table}
\caption{Numerical parameters of N-body simulation}
\begin{tabular}{p{0.25in}p{0.05in}p{0.05in}p{0.05in}p{0.05in}p{0.05in}p{0.4in}p{0.1in}p{0.4in}p{0.4in}} 
\hline 
Run & $\Omega_{0}$ & $\Lambda$ & $h$ & $\Gamma$ & $\sigma _{8}$ & \makecell{L\\(Mpc/h)} & $N$ & \makecell{$m_{p}$\\$M_{\odot}/h$} & \makecell{$l_{soft}$\\(Kpc/h)} \\ 
\hline 
SCDM1 & 1.0 & 0.0 & 0.5 & 0.5 & 0.51 & \centering 239.5 & $256^{3}$ & 2.27$\times 10^{11}$ & \makecell{\centering 36} \\ 
\hline 
\end{tabular}
\label{tab:1}
\end{table}

\section{Real-space inverse mass cascade of SG-CFD }
\label{sec:3}
The real-space inverse mass cascade in dark matter flow has been previously studied \citep{Xu:2021-Inverse-mass-cascade-mass-function}. This section briefly reviews the key findings. We rely on mass flux function to quantify the net transfer of mass from all halos below a given mass scale $m_{h}$ to all halos above that scale at any time \textit{t} (or equivalently, scale factor \textit{a} and redshift \textit{z}). The mass flux function $\Pi _{m} \left(m_{h} ,a\right)$ is defined as
\begin{equation}
\label{ZEqnNum400994} 
\Pi _{m} \left(m_{h} ,a\right)=-\frac{\partial }{\partial t} \left[M_{h} \left(a\right)\int _{m_{h} }^{\infty }f_{M} \left(m,m_{h}^{*} \right) dm\right].      
\end{equation} 
\noindent Here $M_{h} \left(a\right)$ is the total mass in all halos that increases with time, i.e. the total mass of halo sub-system. By contrast, the total mass $M_{o} \left(a\right)$ of out-of-halo sub-system includes all collisionless particles that do not belong to any halos. 

The halo mass function $f_{M} \left(m_{h} ,m_{h}^{*} \left(a\right)\right)$ is the probability distribution of total halo mass $M_{h} \left(a\right)$ with respect to mass $m_{h}$ (or to the number of particles in halo $n_{p} ={m_{h}/m_{p} } $), where $m_{p}$ is the mass of a single particle (or single merger, the mass resolution of \textit{N}-body simulation). The characteristic mass scale $m_{h}^{*} \left(a\right)$ (or $n_{p}^{*} ={m_{h}^{*}/m_{p} } $) gives the size of typical halos formed at time \textit{t} and increases with time. The mass flux function $\Pi _{m} $ should be independent of halo size $m_{h} $ for halo groups smaller than $m_{h}^{*} \left(a\right)$ (i.e. in mass propagation range), where mass flux function reduces to
\begin{equation}
\varepsilon _{m} \left(a\right)=\Pi _{m} \left(m_{h} ,a\right)\quad \textrm{for} \quad m_{h} \ll m_{h}^{*},      
\label{ZEqnNum461250}
\end{equation}
\noindent while flux function $\Pi _{m} $ is mass-scale dependent in mass deposition range with $m_{h} \gg m_{h}^{*}$.

The constant mass flux (or mass dissipation rate $\varepsilon_{m}$) propagates mass from the smallest mass scale to the characteristic scale ($0\ll m_{h} <m_{h}^{*} $) in mass propagation range. Here the total mass of a halo group of size $m_{h}$ is given by,
\begin{equation} 
\label{ZEqnNum444911} 
m_{g} \left(m_{h} ,a\right)=n_{h} m_{h} =M_{h} \left(a\right)f_{M} \left(m_{h} ,m_{h}^{*} \right)m_{p} ,       
\end{equation} 
where $n_{h} $ is the number of halos in a halo group including all halos with the same mass $m_{h}$.

The real-space mass transfer function $T_{m} $ can be defined as the derivative of mass flux function with respect to $m_{h}$, 
\begin{equation} 
\label{ZEqnNum980409}
\begin{split}
T_{m} \left(m_{h} ,a\right)&=\frac{\partial \Pi _{m} \left(m_{h} ,a\right)}{\partial m_{h} }\\
&=\frac{\partial \left[M_{h} \left(a\right)f_{M} \left(m_{h} ,m_{h}^{*} \right)\right]}{\partial t}
=\frac{\partial m_{g} \left(m_{h} ,a\right)}{m_{p} \partial t},
\end{split}
\end{equation} 
which quantifies the rate of change of group mass $m_{g}(m_{h} ,a)$. We can express the mass flux function as the integral of mass transfer function, i.e.
\begin{equation} 
\label{eq:7} 
\Pi _{m} \left(m_{h} ,a\right)=\int _{0}^{m_{h} }T_{m} \left(m,a\right)dm ,        
\end{equation} 
where $T_{m}(m,a)$ clearly represents the rate of mass transfer from halo groups with mass \textit{m} (below $m_{h}$) to halo groups with mass above $m_{h}$. The integral of $T_{m} $ is the total rate of mass transfer $\Pi _{m} $. For mass propagation range (using Eqs. \eqref{ZEqnNum461250} and \eqref{ZEqnNum980409}), 
\begin{equation}
T_{m} \left(m_{h} ,a\right)=0\quad \textrm{and} \quad \frac{\partial m_{g} \left(m_{h} ,a\right)}{\partial t} =0\quad \textrm{for} \quad m_{h} <m_{h}^{*},   
\label{ZEqnNum690015}
\end{equation}

\noindent where halo group mass $m_{g} $ is time-invariant. The mass transfer function $T_{m} \left(m_{h} ,a\right)$ describes the removal of mass from a small scale and the deposition of mass at a large scale, where $T_{m} \left(m_{h} ,a\right)>0$ when $m_{h} >m_{h}^{*} $ indicates that mass is deposited to grow halos above the characteristic size $m_{h}^{*} $. 

The scale-independent mass flux $\varepsilon _{m} $ leads to a constant group mass $m_{g} \left(m_{h} \right)\equiv m_{g} \left(m_{h} ,a\right)$ that reaches a steady state in the propagation range (Eq. \eqref{ZEqnNum690015}). Mass injected at the smallest mass scale is simply propagated through all halo groups below the characteristic size $m_{h}^{*} $ and consumed (deposited) to grow halos above the characteristic mass $m_{h}^{*} $. Mass of halo groups below $m_{h}^{*} $ does not grow with time. The mass scale $m_{h}^{*} $ increases with time and extends the propagation range to larger mass scale. Since mass flux $\varepsilon _{m} \left(a\right)$ is independent of halo mass for $m_{h} <m_{h}^{*}$, the mass flux at the smallest scale reads (Eq. \eqref{ZEqnNum400994} with $m_{h} =0$)
\begin{equation}
\varepsilon _{m} \left(a\right)=\Pi _{m} \left(0,a\right)=-\frac{\partial M_{h} \left(a\right)}{\partial t} =-\left(\frac{3}{2} -\tau _{0} \right)M_{h} \left(a\right)H, 
\label{ZEqnNum512327}
\end{equation}
where the total halo mass $M_{h} \left(a\right)\sim a^{{3/2} -\tau _{0} } $ \citep[see] [Table 2]{Xu:2021-Inverse-mass-cascade-mass-function}. Here $\tau _{0}$ is a mass cascade parameter.

\begin{figure}
\includegraphics*[width=\columnwidth]{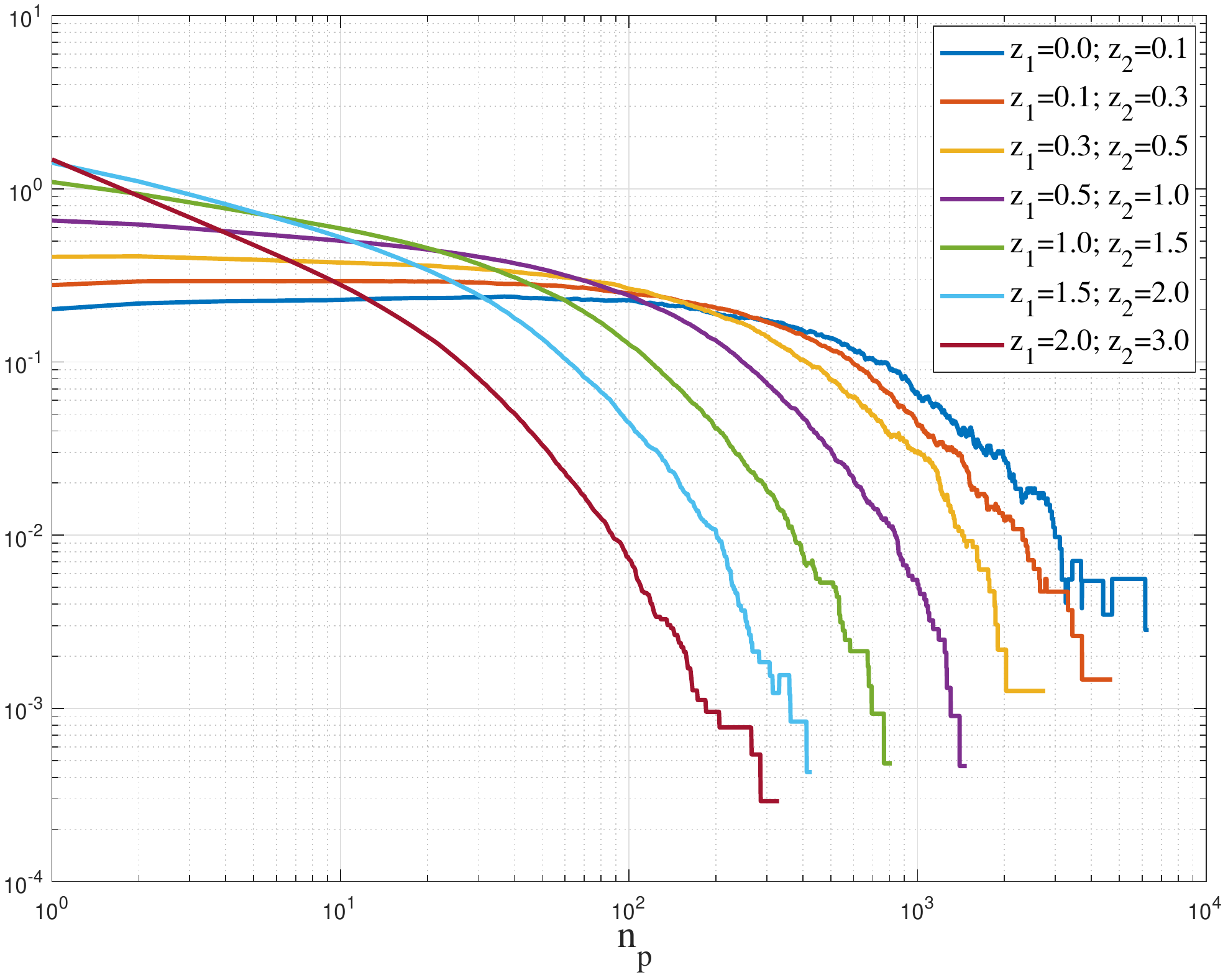}
\caption{Mass flux function $-\Pi _{m} \left(m,a\right)$ (normalized by ${Nm_{p}/t_{0} } $ and $t_{0} $ is the current time) computed from simulation results using halo group mass $m_{g} \left(a\right)$ at two different redshifts \textit{z} (Eq. \eqref{ZEqnNum400994}). A scale-independent mass flux $\varepsilon _{m} \left(a\right)$ can be found for propagation range. The negative mass flux $\varepsilon _{m} \left(a\right)$ clearly indicates an inverse mass cascade from small to large scales. The mass propagation range (with scale-independent $\varepsilon _{m} \left(a\right)$) is formed at $z=0.3$ and extends to larger mass scale with time.}
\label{fig:1}
\end{figure}

Figure \ref{fig:1} plots the mass flux function $-\Pi _{m} \left(m,a\right)$ computed from a \textit{N}-body simulation using halo group mass $m_{g} \left(a\right)$ at two different redshifts \textit{z} (Eq. \eqref{ZEqnNum400994}). A scale-independent mass flux $\varepsilon _{m} \left(a\right)$ can be clearly identified in the propagation range ($m_{h} <m_{h}^{*} $). The negative mass flux function $\varepsilon _{m} \left(a\right)<0$ indicates an inverse mass cascade from small to large scales. The propagation range is clearly formed at around $z=0.3$ and extends to larger mass scale with time.

Let time scale $\tau _{h} \left(m_{h} ,a\right)$ be the average waiting time for a single merging event in a halo group of mass $m_{h} $. The scale-independent rate of mass transfer from scale below $m_{h} $ to above $m_{h}$ is 
\begin{equation}
\varepsilon _{m} \left(a\right)=-{m_{h}/\tau _{h} \left(m_{h} ,a\right)} \quad \textrm{for} \quad m_{h} \ll m_{h}^{*}.     
\label{ZEqnNum379703}
\end{equation}

\noindent Because halo interactions are local in mass space, we can assume the dominant merging events in an infinitesimal time involves merging between a halo and a single merger. This is a fundamental step in mass cascade and explored by a two-body collapse model, i.e. TMCM model \citep{Xu:2021-A-non-radial-two-body-collapse}, that can provide more insights than a simple spherical collapse model \citep{Gunn:1977-Massive-Galactic-Halos--1--For}. The average waiting time $\tau _{g} =n_{h} \tau _{h} $, i.e. halo lifespan \citep[see][Eq. (5)]{Xu:2021-Inverse-mass-cascade-mass-function} for a halo of mass $m_{h}$,  is $\tau _{g} \propto a^{\tau _{0} } m_{h}^{-\lambda } $ \citep[see][Eq. (45)]{Xu:2021-Inverse-mass-cascade-mass-function}, where $\tau _{0} \approx -1$ is usually assumed and $\lambda $ is a halo geometry parameter. 

With $\lambda ={2/3}$, mass of a typical halo $m_{h}^{L} \left(t\right)\sim t$ for halos with a deterministic waiting time or lifespan $\tau _{g}$ \citep[see][Eq. (9)]{Xu:2021-Inverse-mass-cascade-mass-function}. The actual halo lifespan can be a random variable. Now let's assume a typical halo of mass $m_{h}^{L} \left(t\right)$ that is constantly growing with the waiting time exactly to be $\tau _{g} $ for every single merging during its entire mass accretion history. The total halo mass $M_{h} \left(a\right)$ can be related to the mass of that typical halo $m_{h}^{L}$ \citep[see][Eq. (53)]{Xu:2021-Inverse-mass-cascade-mass-function}, 
\begin{equation} 
\label{ZEqnNum391224} 
M_{h} \left(a\right)=\frac{1}{1-\lambda } m_{h}^{L} n_{h}^{L} n_{p}^{L} =\frac{1}{\beta _{0} } m_{h}^{*} n_{h}^{*} n_{p}^{*} ,        
\end{equation} 
where $\beta _{0} $ is a numerical constant for mass function $f_{M} \left(m_{h} ,m_{h}^{*} \left(a\right)\right)$. The scale-independent mass flux $\varepsilon _{m} $ can be eventually expressed as \citep[see][Eq. (54)]{Xu:2021-Inverse-mass-cascade-mass-function}
\begin{equation}
\label{ZEqnNum450896} 
\varepsilon _{m} \left(a\right)=-\frac{dm_{h}^{L} }{dt} n_{h}^{L} n_{p}^{L} =-\frac{1}{\left(1-\lambda \right)} \frac{d\left(m_{h}^{L} n_{h}^{L} n_{p}^{L} \right)}{dt},    
\end{equation} 
where $n_{h}^{L} $ is the number of that typical halos in the halo group of mass $m_{h}^{L} $ and $n_{p}^{L} $ is the number of particles in that typical halo. Here $n_{h}^{L} \sim \small(m_{h}^{L} \small)^{-(1+\lambda)}$ that decreases with halo mass $m_{h}^{L}$.

In hydrodynamic turbulence, the vortex stretching is generally considered as the major mechanism to facilitate the energy transfer from large to small length scales (Eq. \eqref{ZEqnNum426779}). Vortex stretches into smaller scales with kinetic energy transferred to smaller scales simultaneously. In dark matter flow (SG-CFD), halos are the building block in SG-CFD. Equation \eqref{ZEqnNum391224} tells us that entire halo sub-system can be equivalently treated as consisting of only typical halos of mass $m_{h}^{L}$ with an equivalent number of halos $n_{h}^{L} n_{p}^{L} $. Equation \eqref{ZEqnNum450896} quantifies the mass flux $\varepsilon _{m} $ in terms of the growth of typical halos. Halos constantly merges with single mergers from out-of-halo sub-system, which facilitates the inverse mass cascade from small to large mass scales. It was also shown that \citep[see][Eq. (50)]{Xu:2021-Inverse-mass-cascade-mass-function} mass of typical halos and total halo mass scales with \textit{a}, 
\begin{equation}
\begin{split}
&m_{h}^{L} \left(a\right)\sim a^{\lambda _{m} } =a^{\frac{{3/2} -\tau _{0} }{1-\lambda } }\\ 
&\textrm{and}\\
&M_{h} \left(a\right)\sim a^{\left(1-\lambda \right)\lambda _{m} } \quad \textrm{with} \quad \lambda _{m} =\frac{{3/2} -\tau _{0} }{1-\lambda }. 
\end{split}
\label{ZEqnNum440765}
\end{equation}

\noindent If $\tau _{0} =1$ and $\lambda ={2/3} $ (i.e. $\lambda _{m} ={3/2} $), we should have $m_{h}^{L} \left(a\right)\sim a^{{3/2} } \sim t$, $n_{h}^{L} n_{p}^{L} \sim \left(m_{h}^{L} \right)^{-\lambda } \sim a^{-1} $ and $M_{h}^{} \left(a\right)\sim a^{{1/2} } $. Typical halos grow at a constant rate with time while the equivalent number of typical halos $\left(n_{h}^{L} n_{p}^{L} \right)$ decreases as $a^{-1} $ such that mass flux function $\varepsilon _{m} \left(a\right)\sim a^{-1} $ (Eq. \eqref{ZEqnNum450896}). 

Just like the mass cascade, energy cascade can be facilitated by the growth of those typical halos via merging with single mergers. With halos growing from small to large mass scales, the total kinetic energy of that halo increases while potential energy decreases such that energy is cascaded across mass scales. This qualitative picture hints an inverse kinetic energy cascade and a direct potential energy cascade. Quantitative analysis is provided in the next section.

\section{Real-space energy cascade of SG-CFD}
\label{sec:4}
\subsection{Inverse cascade of halo kinetic energy}
\label{sec:4.1}
Individual halos are characterized by halo size (number of particles $n_{p}$ or halo mass $m_{h}$), one-dimensional halo virial dispersion ($\sigma _{v}^{2} $), and halo velocity ($\boldsymbol{\mathrm{u}}_{h} $) (the mean velocity of all particles in that halo). Particle velocity $\boldsymbol{\mathrm{u}}_{p} $ can be decomposed into 
\begin{equation} 
\label{ZEqnNum502045} 
\boldsymbol{\mathrm{u}}_{p} =\boldsymbol{\mathrm{u}}_{h} +\boldsymbol{\mathrm{u}}_{p}^{'} ,           
\end{equation} 
where $\boldsymbol{\mathrm{u}}_{p}^{'} $ is the fluctuation of particle velocity around halo velocity $\boldsymbol{\mathrm{u}}_{h} $. The virial dispersion $\sigma _{vh}^{2} $ is the dispersion of fluctuation $\boldsymbol{\mathrm{u}}_{p}^{'}$
\begin{equation} 
\label{ZEqnNum627513} 
\sigma _{vh}^{2} =var\left(\boldsymbol{\mathrm{u}}_{p}^{'x} \right)=var\left(\boldsymbol{\mathrm{u}}_{p}^{'y} \right)=var\left(\boldsymbol{\mathrm{u}}_{p}^{'z} \right),        
\end{equation} 
i.e. the variance of velocity fluctuation for all particles in the same halo. Virial dispersion $\sigma _{vh}^{2} $  represents the mean kinetic energy of particles or the temperature of that halo. 

The halo velocity $\boldsymbol{\mathrm{u}}_{\boldsymbol{\mathrm{h}}} =\left\langle \boldsymbol{\mathrm{u}}_{\boldsymbol{\mathrm{p}}}^{} \right\rangle _{h} $ is the mean velocity of all particles in the same halo, where $\left\langle \right\rangle _{h} $ stands for the average for all particles in the same halo. All halos identified in the system can be grouped into halo groups of the same size. Halo groups are characterized by the size of halos in that group ($n_{p} $ or $m_{h} $), mean halo virial dispersion ($\sigma _{v}^{2} $), and one-dimensional halo velocity dispersion ($\sigma _{h}^{2} $) that is defined as the dispersion (variance) of halo velocity $\boldsymbol{\mathrm{u}}_{h} $ for all halos in the same group, 
\begin{equation}
\sigma _{v}^{2} =\left\langle \sigma _{vh}^{2} \right\rangle _{g}, \quad \sigma _{h}^{2} =var\left(\boldsymbol{\mathrm{u}}_{h}^{x} \right)=var\left(\boldsymbol{\mathrm{u}}_{h}^{y} \right)=var\left(\boldsymbol{\mathrm{u}}_{h}^{z} \right).     
\label{eq:16}
\end{equation}

\noindent The halo velocity dispersion $\sigma _{h}^{2}$ represents the mean kinetic energy of halos or the temperature of halo group, while mean virial dispersion $\sigma _{v}^{2}$ is the mean halo temperature with $\left\langle \right\rangle _{g} $ for the average for all halos in the same group. Note that the statistics defined at the halo group level is the statistics over many replicas (possible states) of halos with the same mass $m_{h} $. Dispersion of particle velocity of all particles in the same group can be decomposed accordingly,
\begin{equation} 
\label{ZEqnNum425919} 
\sigma ^{2} =\sigma _{h}^{2} +\sigma _{v}^{2} .           
\end{equation} 

\begin{figure}
\includegraphics*[width=\columnwidth]{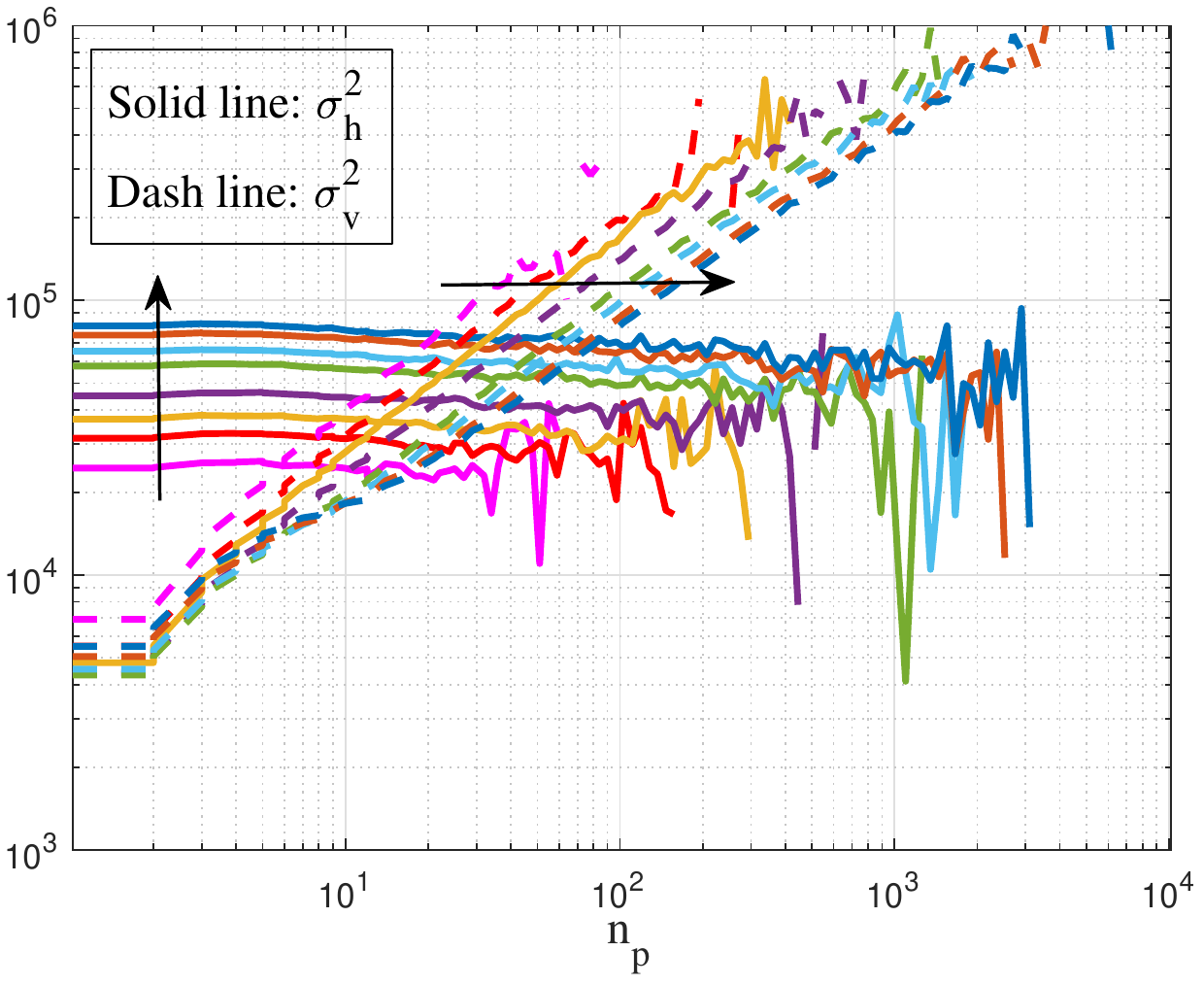}
\caption{The variation of halo velocity dispersion $\sigma _{h}^{2} $ ($(Km/s)^2$, solid lines) and halo virial dispersion $\sigma _{v}^{2} $ ($(Km/s)^2$, dash lines) with halo group size $n_{p} $ for different redshifts z = 0, 0.1, 0.3, 0.5, 1.0, 1.5, 2.0, and 3.0. Arrows points to the direction of the increasing time or decreasing redshift \textit{z}. The halo velocity dispersion $\sigma _{h}^{2} $ is relatively independent of halo mass and increases with scale factor \textit{a}, while halo virial dispersion $\sigma _{v}^{2} $ increases with halo mass and decreases with \textit{a} for a given mass $m_h$ or size $n_p$.}
\label{fig:2}
\end{figure}

Figure \ref{fig:2} plots the variation of halo velocity dispersion $\sigma _{h}^{2} \left(m_{h} ,a\right)$ and virial dispersion $\sigma _{v}^{2} \left(m_{h} ,a\right)$ with group size $n_{p} $ at different redshifts \textit{z}. Halo velocity dispersion $\sigma _{h}^{2} $ increases with scale factor \textit{a}, while halo virial dispersion $\sigma _{v}^{2} $ decreases with \textit{a} for a given size $m_{h} $. The approximate fitting relations are 
\begin{equation} 
\label{ZEqnNum259105} 
\sigma _{h}^{2} \left(a\right)\approx \left\langle \sigma _{h}^{2} \right\rangle \left(a\right)=\beta _{\sigma h} au_{0}^{2} ,         
\end{equation} 
and
\begin{equation} 
\label{ZEqnNum222387} 
\sigma _{v}^{2} \left(m_{h} ,a\right)\approx \beta _{\sigma v} a^{-1} u_{0}^{2} \left({m_{h}/m_{p} } \right)^{{2/3} } , 
\end{equation} 
with $\beta _{\sigma h} =0.57$ and $\beta _{\sigma v} =0.03$ (see e.g. \cite{Bryan:1998-Statistical-properties-of-X-ra} for other fitting formula), where $m_{p} =2.27\times 10^{11} {M_{\odot }/h} $ is mass resolution in Table \ref{tab:1}. Here $u_{0}=354.61km/s$ is the one-dimensional velocity dispersion of entire system at present epoch (\textit{z}=0). The virial dispersion of small halos is relatively time-invariant and less dependent on scale factor \textit{a} due to their compact stable core structures.  

Based on this description, the kinetic energy of a given particle can be separated into two contributions, i.e. halo kinetic energy $\sigma _{h}^{2} $ from the random motion of halos and virial kinetic energy $\sigma _{v}^{2} $ from the particle motion in halos. The halo kinetic energy is dominant for small halos with $\sigma _{h}^{2} \gg \sigma _{v}^{2} $, while the virial kinetic energy is dominant for large halos $\sigma _{v}^{2} \gg \sigma _{h}^{2} $. Since mass cascade is facilitated by halo merging with single mergers, it is reasonable to expect that the energy cascade follows a similar way as mass cascade, where the kinetic energy is injected at small scales, propagated through the mass propagation range, and deposited to grow the kinetic energy of large halos in the mass deposition range. 

Here we first focus on the halo kinetic energy $\sigma _{h}^{2} $ and introduce a kinetic energy flux function that quantifies the net transfer of halo kinetic energy $\sigma _{h}^{2} $ from all halos smaller than $m_{h}$ to all halos greater than $m_{h}$. Just like Eq. \eqref{ZEqnNum400994}, kinetic energy flux $\Pi _{kh} \left(m_{h} ,a\right)$ can be defined as
\begin{equation} 
\label{ZEqnNum683782} 
\begin{split}
&\Pi _{kh} \left(m_{h} ,a\right)=\underbrace{M_{h} \left(a\right)\int _{m_{h} }^{\infty }f_{M} \left(m,m_{h}^{*} \right) \frac{\partial \sigma _{h}^{2} }{\partial t} dm}_{1}\\
&\quad\quad-\underbrace{\frac{\partial }{\partial t} \left[M_{h} \left(a\right)\int _{m_{h} }^{\infty }f_{M} \left(m,m_{h}^{*} \right) \sigma _{h}^{2} \left(m,a\right)dm\right]}_{2},
\end{split}
\end{equation} 
where term 2 is the rate of change of total halo kinetic energy contained in halos greater than $m_{h} $. While the total kinetic energy of a forced steady turbulence is conserved and does not varying with time, the total kinetic energy in SG-CFD does increase with time due to exchange with potential energy (more discussion in Section \ref{sec:4.5}). 

Comparing with the mass flux $\Pi _{m} \left(m_{h} ,a\right)$ in Eq. \eqref{ZEqnNum400994}, the extra term (term 1) of $\Pi _{kh} \left(m_{h} ,a\right)$ accounts for the change of halo kinetic energy contained in halos greater than $m_{h} $ that is simply due to the time-variation of $\sigma _{h}^{2} \left(m_{h} ,a\right)$, which does not come from the energy cascade and should be excluded. The energy flux after combining two terms together reads 
\begin{equation} 
\label{ZEqnNum490431} 
\begin{split}
\Pi _{kh} \left(m_{h} ,a\right)&=-\int _{m_{h} }^{\infty }\frac{\partial }{\partial t} \left[M_{h} \left(a\right)f_{M} \left(m,m_{h}^{*} \right)\right] \sigma _{h}^{2} \left(m,a\right)dm\\
&=-\int _{m_{h} }^{\infty }T_{m} \left(m,a\right) \sigma _{h}^{2} \left(m,a\right)dm.
\end{split}
\end{equation} 
The energy transfer function can be introduced similar to Eq. \eqref{ZEqnNum980409},
\begin{equation} 
\label{ZEqnNum928831} 
\begin{split}
T_{kh} \left(m_{h} ,a\right)=\frac{\partial \Pi _{kh} }{\partial m_{h} }
&=\frac{\partial }{\partial t} \left[M_{h} \left(a\right)f_{M} \left(m_{h} ,m_{h}^{*} \right)\right]\sigma _{h}^{2} \left(m_{h} ,a\right)\\
&=T_{m} \left(m_{h} ,a\right)\sigma _{h}^{2} \left(m_{h} ,a\right), 
\end{split}
\end{equation} 
which is proportional to the mass transfer function $T_{m} \left(m_{h} ,a\right)$ for mass cascade. Clearly, the energy cascade is intimately related to the mass cascade, while mass cascade does not exist in turbulence. 

To compute the rate of energy production at the smallest scale, let's introduce the mean rate of energy change
\begin{equation} 
\label{ZEqnNum657027} 
\left\langle \frac{\partial \sigma _{h}^{2} }{\partial t} \right\rangle =\int _{0}^{\infty }f_{M} \left(m_{h} ,m_{h}^{*} \right) \frac{\partial \sigma _{h}^{2} }{\partial t} dm_{h}  
\end{equation} 
and the mean (specific) halo kinetic energy 
\begin{equation} 
\label{eq:24} 
\left\langle \sigma _{h}^{2} \right\rangle =\int _{0}^{\infty }f_{M} \left(m_{h} ,m_{h}^{*} \right) \sigma _{h}^{2} \left(m_{h} ,a\right)dm_{h} .        
\end{equation} 
From Eq. \eqref{ZEqnNum683782}, the rate of energy production at the smallest scale is 
\begin{equation}
\label{ZEqnNum798916} 
\Pi _{kh} \left(m_{h} \to 0,a\right)=M_{h} \left(a\right)\left\langle \frac{\partial \sigma _{h}^{2} }{\partial t} \right\rangle -\frac{\partial }{\partial t} \left[M_{h} \left(a\right)\left\langle \sigma _{h}^{2} \right\rangle \right].     
\end{equation} 
These expressions can be simplified for a scale-independent halo velocity dispersion, where  $\sigma _{h}^{2} \left(m_{h} ,a\right)\equiv \left\langle \sigma _{h}^{2} \right\rangle \left(a\right)$. The flux function of $\sigma _{h}^{2} $ reads (from Eqs. \eqref{ZEqnNum259105} and \eqref{ZEqnNum490431})
\begin{equation} 
\label{eq:26} 
\Pi _{kh} \left(m_{h} ,a\right)\approx \Pi _{m} \left(m_{h} ,a\right)\left\langle \sigma _{h}^{2} \right\rangle ,        
\end{equation} 
where the energy flux $\Pi _{kh} $ is proportional to the mass flux $\Pi _{m} $. The rate of energy production at the smallest scale should be (from Eqs. \eqref{ZEqnNum259105} and \eqref{ZEqnNum798916}),
\begin{equation} 
\label{ZEqnNum894335} 
\varepsilon _{kh} =\Pi _{kh} \left(0,a\right)=\varepsilon _{m} \left\langle \sigma _{h}^{2} \right\rangle =-\left(\frac{3}{2} -\tau _{0} \right)M_{h} \left(a\right)H\left\langle \sigma _{h}^{2} \right\rangle  
\end{equation} 
For $\tau _{0} =1$, we have $M_{h} \left(a\right)\propto a^{{1/2} } $ and $\sigma _{h}^{2} \sim a$ from Eq. \eqref{ZEqnNum259105}. The rate of energy production $\varepsilon _{kh}$ should be independent of time and reaches a steady state. Equivalently, the total halo kinetic energy in halo sub-system should be proportional to time \textit{t} (detail discussion also in Fig. \ref{fig:10}). 

At statistically steady state, the mass transfer function $T_{m} \left(m_{h} ,a\right)\approx 0$ for propagation range (Eq. \eqref{ZEqnNum690015}). The kinetic energy flux function $\Pi _{kh} $ across different size of halos should be independent of halo size $m_{h}$, where we have
\begin{equation}
T_{kh} \left(m_{h} ,a\right)=0,\quad \varepsilon _{kh} \left(a\right)=\Pi _{kh} \left(m_{h} ,a\right) \quad \textrm{for} \quad m_{h} <m_{h}^{*},  
\label{eq:28}
\end{equation}
We can similarly define the total halo kinetic energy in a halo group of same size halos (similar to Eq. \eqref{ZEqnNum444911} for $m_g$), 
\begin{equation} 
\label{eq:29} 
\begin{split}
\sigma _{hg}^{2} \left(m_{h} ,a\right)&=M_{h} \left(a\right)f_{M} \left(m_{h} ,m_{h}^{*} \right)\sigma _{h}^{2} \left(m_{h} ,a\right)\\
&=\sigma _{h}^{2} \left(m_{h} ,a\right){m_{g} \left(m_{h} ,a\right)/m_{p} }.
\end{split}
\end{equation} 
the rate of change of $\sigma _{hg}^{2}$ is,
\begin{equation} 
\label{eq:30} 
\frac{\partial \sigma _{hg}^{2} }{\partial t} =\frac{\partial m_{g} \left(m_{h} ,a\right)}{\partial t} \frac{\sigma _{h}^{2} \left(m_{h} ,a\right)}{m_{p} } +\frac{m_{g} }{m_{p} } \frac{\partial \sigma _{h}^{2} \left(m_{h} ,a\right)}{\partial t} .      
\end{equation} 
A direct result of a scale-independent mass flux through halos of different mass scales is a time-invariant group mass $m_{g} $ such that,
\begin{equation}
\frac{\partial \sigma _{hg}^{2} }{\partial t} =\frac{m_{g} }{m_{p} } \frac{\partial \sigma _{h}^{2} \left(m_{h} ,a\right)}{\partial t} \quad \textrm{for} \quad m_{h} \ll m_{h}^{*},       
\label{eq:31}
\end{equation}

\noindent i.e. the change of $\sigma _{hg}^{2} $ in propagation range is only due to the time variation of $\sigma _{h}^{2} $. The energy cascade simply propagates (not changes) halo kinetic energy through halos in propagation range.

\begin{figure}
\includegraphics*[width=\columnwidth]{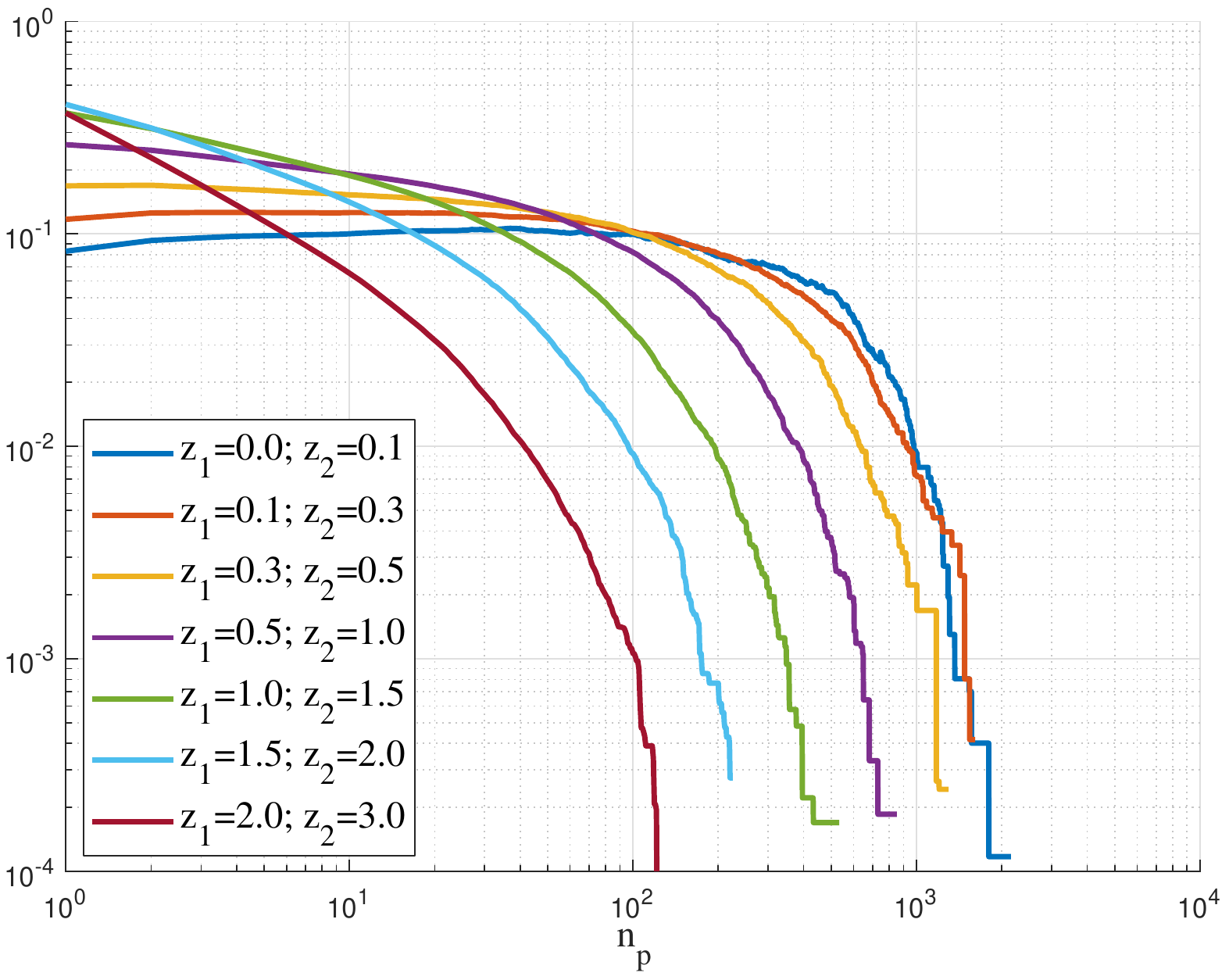}
\caption{The variation of energy flux function $-\Pi _{kh} $ for halo kinetic energy $\sigma _{h}^{2} $ with size $n_{p} $ of halo groups. The flux function $\Pi _{kh} <0$ (inverse cascade from small to large scales) is normalized by ${Nm_{p} u_{0}^{2}/t_{0} } $. The flux function $\Pi _{kh} $ is computed from simulation results at two different redshifts $z_{1} $ and $z_{2} $. A scale-independent flux function $\varepsilon _{kh} =\Pi _{kh} \left(m_{h} ,a\right)$ can be identified for the mass propagation range with $m_{h} <m_{h}^{*} $.}
\label{fig:3}
\end{figure}

Figure \ref{fig:3} plots the variation of flux function $-\Pi _{kh} \left(m_{h} ,a\right)$ of halo kinetic energy $\sigma _{h}^{2} $ with halo group size $n_{p}$. The flux function $\Pi _{kh} $ is negative indicating an inverse cascade from small to large mass scales. The flux function $\Pi _{kh} $ is numerically computed from simulation results at two different redshifts $z_{1} $ and $z_{2} $ using Eqs. \eqref{ZEqnNum683782} or \eqref{ZEqnNum490431} and dataset \citep{Xu:2022-Dark_matter-flow-dataset-part1,Xu:2022-Dark_matter-flow-dataset-part2}. A scale-independent flux function $\varepsilon _{kh} $ can be identified for the mass propagation range with $m_{h} <m_{h}^{*} $.

\subsection{General formulation for cascade of a specific quantity}
\label{sec:4.2}
In principle, the flux function for any specific quantity $V_{s} $ (an extensive quantity per unit mass) can be similarly derived. For example, let us assume a generic specific variable $V_{s} \left(m_{h} ,a\right)$ that can be expressed in a general form via separation of variables,
\begin{equation} 
\label{ZEqnNum130199} 
V_{s} \left(m_{h} ,a\right)=a^{s} F_{v} \left(m_{h} \right),         
\end{equation} 
where \textit{s} is an arbitrary exponent. The flux function of $V_{s} $ can be written as (From Eq. \eqref{ZEqnNum490431}),
\begin{equation} 
\label{eq:33} 
\Pi _{vs} \left(m_{h} ,a\right)=-\int _{m_{h} }^{\infty }T_{m} \left(m,a\right) V_{s} \left(m,a\right)dm,       
\end{equation} 
where the transfer function $T_{vs} $ of quantity $V_{s} $ is proportional to the mass transfer function $T_{m} $,
\begin{equation} 
\label{eq:34} 
T_{vs} \left(m_{h} ,a\right)=T_{m} \left(m_{h} ,a\right)V_{s} \left(m_{h} ,a\right).        
\end{equation} 
The rate of production for $V_{s}$ reads (from Eqs. \eqref{ZEqnNum683782} and \eqref{ZEqnNum798916})
\begin{equation} 
\label{ZEqnNum738820} 
\varepsilon _{vs} =\Pi _{vs} \left(m_{h} \to 0,a\right)=M_{h} \left(a\right)\left\langle \frac{\partial V_{s} }{\partial t} \right\rangle -\frac{\partial }{\partial t} \left[M_{h} \left(a\right)\left\langle V_{s} \right\rangle \right],     
\end{equation} 
where the mean rate of change for quantity $V_{s} $
\begin{equation} 
\label{ZEqnNum710390} 
\left\langle \frac{\partial V_{s} }{\partial t} \right\rangle =\int _{0}^{\infty }f_{M} \left(m_{h} ,m_{h}^{*} \right) \frac{\partial V_{s} }{\partial t} dm_{h}  
\end{equation} 
and the mean (specific) variable $V_{s}$ reads
\begin{equation} 
\label{ZEqnNum317378} 
\left\langle V_{s} \right\rangle =\int _{0}^{\infty }f_{M} \left(m_{h} ,m_{h}^{*} \right) V_{s} dm_{h} .         
\end{equation} 
With the general expression for $V_{s} $ in Eq. \eqref{ZEqnNum130199}, we should have
\begin{equation} 
\label{eq:38} 
\left\langle \frac{\partial V_{s} }{\partial t} \right\rangle =sH\left\langle V_{s} \right\rangle  
\end{equation} 
from definition \eqref{ZEqnNum710390} and \eqref{ZEqnNum317378}. The rate of production for $V_{s} $ at the smallest scale now should be (Eq. \eqref{ZEqnNum738820})
\begin{equation} 
\label{ZEqnNum593887} 
\varepsilon _{vs} =-M_{h} \left(a\right)H\left\langle V_{s} \right\rangle \left[-s+\frac{\partial \ln \left[M_{h} \left(a\right)\left\langle V_{s} \right\rangle \right]}{\partial \ln a} \right],      
\end{equation} 
or equivalently (using Eq. \eqref{ZEqnNum512327}), 
\begin{equation} 
\label{ZEqnNum802989} 
\begin{split}
\varepsilon _{vs}&=\frac{\varepsilon _{m} \left\langle V_{s} \right\rangle }{{3/2} -\tau _{0} } \left[-s+\frac{\partial \ln \left[M_{h} \left(a\right)\left\langle V_{s} \right\rangle \right]}{\partial \ln a} \right]\\
&=\frac{\varepsilon _{m} \left\langle V_{s} \right\rangle }{{3/2} -\tau _{0} } \left[\frac{\partial \ln \left\langle V_{s} \right\rangle }{\partial \ln a} -s+\frac{3}{2} -\tau _{0} \right].
\end{split}
\end{equation} 
Clearly, the flux function of any specific quantity is proportional to the flux of mass transfer and the mean of that quantity, i.e. $\varepsilon _{vs} \sim \varepsilon _{m} \left\langle V_{s} \right\rangle $. A scale-independent mass flux $\varepsilon _{m} $  will lead to a scale-independent flux $\varepsilon _{vs} $ for any specific quantity $V_{s}$. By setting $V_{s} =\sigma _{h}^{2} $ and $s=1$, the flux function of halo kinetic energy $\varepsilon _{kh} $ (Eq. \eqref{ZEqnNum894335}) can be easily recovered from Eq. \eqref{ZEqnNum802989}.

Since mass cascade can be quantitatively described by the growth of typical halos (Eq. \eqref{ZEqnNum450896}), similar treatment can be done for the specific quantity $V_{s} $. If we assume a power law for the mean quantity $\left\langle V_{s} \right\rangle \sim a^{\lambda _{v} } $ and $m_{h}^{L} \sim a^{\lambda _{m} } $ (Eq. \eqref{ZEqnNum440765}), the flux function of quantity $V_{s} $ can be directly written as (from Eqs. \eqref{ZEqnNum802989}, \eqref{ZEqnNum391224} and \eqref{ZEqnNum450896}),
\begin{equation} 
\label{eq:41} 
\varepsilon _{vs} =-\frac{\lambda _{v} -s+{3/2} -\tau _{0} }{\left(\lambda _{m} +\lambda _{v} \right)\left(1-\lambda \right)} \frac{d\left(m_{h}^{L} \left\langle V_{s} \right\rangle \right)}{dt} n_{h}^{L} n_{p}^{L} ,       
\end{equation} 
or
\begin{equation} 
\label{ZEqnNum541860} 
\begin{split}
\varepsilon _{vs}&=\left(\frac{s}{\lambda _{v} +{3/2} -\tau _{0} } -1\right)\frac{d\left(M_{h} \left\langle V_{s} \right\rangle \right)}{dt}\\ &=-\left(\lambda _{v} +{3/2} -\tau _{0} -s\right)M_{h} H\left\langle V_{s} \right\rangle,
\end{split}
\end{equation} 
where the flux function of $V_{s} $ can be conveniently related to the rate of change of total quantity$M_{h} \left\langle V_{s} \right\rangle $ in all halos. It is very often that $\lambda _{v} =1$ such that the flux function $\varepsilon _{vs} $ is constant of time. 

One step further, if the quantity $V_{s}^{L} \equiv V_{s} \left(m_{h}^{L} ,a\right)$ for typical halos of mass $m_{h}^{L} $ can be determined, an equivalent number of typical halos $N_{h}^{L} $ in halo sub-system can be defined as (from Eq. \eqref{ZEqnNum541860}), 
\begin{equation} 
\label{eq:43} 
N_{h}^{L} =\frac{M_{h} \left\langle V_{s} \right\rangle }{m_{h}^{L} V_{s}^{L} } =\frac{n_{h}^{L} n_{p}^{L} }{1-\lambda } \frac{\left\langle V_{s} \right\rangle }{V_{s}^{L} } ,         
\end{equation} 
such that $M_{h} \left\langle V_{s} \right\rangle =N_{h}^{L} m_{h}^{L} V_{s}^{L} $. Now the flux function for the cascade of any quantity $V_{s} $ can be directly related to the mass accretion of the typical halo from Eq. \eqref{ZEqnNum541860}, 
\begin{equation} 
\label{ZEqnNum934449} 
\begin{split}
\varepsilon _{vs} &=-\frac{\lambda _{v} -s+{3/2} -\tau _{0} }{\left(\lambda _{m} +\lambda _{v} -\lambda \lambda _{m} \right)} \frac{d\left(m_{h}^{L} N_{h}^{L} V_{s}^{L} \right)}{dt}\\
&=-\left(\lambda _{v} +{3/2} -\tau _{0} -s\right)Hm_{h}^{L} N_{h}^{L} V_{s}^{L}.    
\end{split}
\end{equation} 
Examples of the specific quantity $V_{s}^{L} $ can be halo specific kinetic/potential energies and radial/rotational kinetic energies from coherent motion in halos. For example, the virial kinetic energy of a typical halo is $\sigma _{v}^{2} \left(m_{h}^{L} ,a\right)\sim a^{-1} \left(m_{h}^{L} \right)^{{2/3} } \sim a^{0} $ (from Eq. \eqref{ZEqnNum222387}) that should be time-variant for $m_{h}^{L} \sim a^{{3/2} } $. The equivalent number of halos $N_{h}^{L} $ is on the order of $\sim 10^{4} $ with both $n_{h}^{L} \sim 100$ and $n_{p}^{L} \sim 100$. Equation \eqref{ZEqnNum934449} can be conveniently used to estimate the flux function of any quantity. A good example is for the rotational kinetic energy discussed in Section \ref{sec:4.6} (Fig. \ref{fig:18}). 

In Hydrodynamic turbulence, the vortex stretching along its axis of rotation facilitates a direct energy cascade from large to small length scales. Similar for SG-CFD, Eq. \eqref{ZEqnNum934449} tells us that the mass accretion of typical halos (the counterpart of vortex) facilitates the cascade of any specific quantity $V_{s} $ from small to large mass scales. 

\subsection{Inverse cascade of virial kinetic energy}
\label{sec:4.3}
Next, the general formulation can be used to formulate the cascade of halo virial energy $\sigma _{v}^{2}$ in Eq. \eqref{eq:16}. The flux function of halo virial energy $\sigma _{v}^{2}$ is first obtained from Eq. \eqref{ZEqnNum490431},  
\begin{equation} 
\label{ZEqnNum258650} 
\Pi _{kv} \left(m_{h} ,a\right)=-\int _{m_{h} }^{\infty }T_{m} \left(m,a\right) \sigma _{v}^{2} \left(m,a\right)dm.
\end{equation} 
Unlike the flux function$\Pi _{kh} \approx \varepsilon _{m} \sigma _{h}^{2} \left(a\right)$ for halo kinetic energy $\sigma _{h}^{2} $ in Eq. \eqref{ZEqnNum894335}, $\Pi _{kv} \ne \varepsilon _{m} \sigma _{v}^{2} \left(m_{h} ,a\right)$ because of the mass-scale dependence of the virial energy $\sigma _{v}^{2} $ (Eq. \eqref{ZEqnNum222387}). The transfer function of virial kinetic energy can be introduced similarly (Eq. \eqref{ZEqnNum928831}),
\begin{equation}
\label{eq:46} 
T_{kv} \left(m_{h} ,a\right)=T_{m} \left(m_{h} ,a\right)\sigma _{v}^{2} \left(m_{h} ,a\right).        
\end{equation} 
The rate of production of halo virial energy at the smallest scale can be obtained from Eq. \eqref{ZEqnNum802989},
\begin{equation} 
\label{ZEqnNum725290} 
\varepsilon _{kv} =\Pi _{kv} \left(0,a\right)=\frac{\varepsilon _{m} \left\langle \sigma _{v}^{2} \right\rangle }{{3/2} -\tau _{0} } \left[1+\frac{\partial \ln \left[M_{h} \left(a\right)\left\langle \sigma _{v}^{2} \right\rangle \right]}{\partial \ln a} \right].     
\end{equation} 
If the mean halo virial kinetic energy $\left\langle \sigma _{v}^{2} \right\rangle \sim a$, i.e. the same scaling as halo kinetic energy $\left\langle \sigma _{h}^{2} \right\rangle $ in Eq. \eqref{ZEqnNum259105} , we will finally have (using Eqs. \eqref{ZEqnNum725290} and \eqref{ZEqnNum450896}),
\begin{equation} 
\label{ZEqnNum316450} 
\varepsilon _{kv} =\left(\frac{7/2-\tau _{0}}{{3/2} -\tau _{0}} \right)\varepsilon _{m} \left\langle \sigma _{v}^{2} \right\rangle =-\left(\frac{7}{2} -\tau _{0} \right)M_{h} \left(a\right)H\left\langle \sigma _{v}^{2} \right\rangle.
\end{equation} 

\begin{figure}
\includegraphics*[width=\columnwidth]{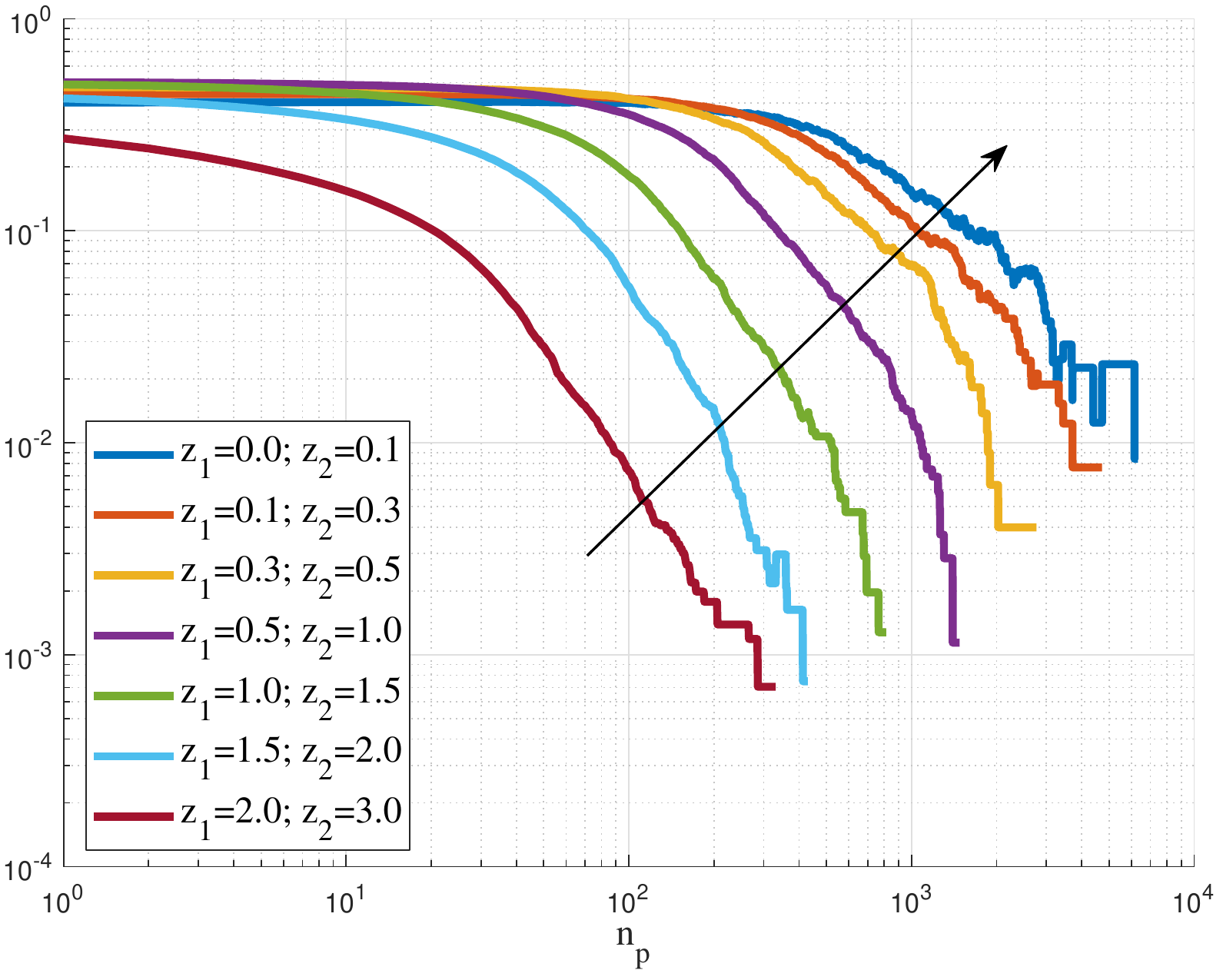}
\caption{The variation of flux function $-\Pi _{kv} \left(m_{h} ,a\right)$ for halo virial energy $\sigma _{v}^{2} \left(m_{h} \right)$ with size $n_{p} $ of halo groups. The flux function $\Pi _{kv} <0$ (inverse cascade from small to large scales) is normalized by ${Nm_{p} u_{0}^{2}/t_{0} } $. The flux function $\Pi _{kv} $ is computed from simulation results at two different redshifts $z_{1} $ and $z_{2} $. A scale-independent flux function $\varepsilon _{kv} $ can be identified for mass propagation range with $m_{h} <m_{h}^{*} $.}
\label{fig:4}
\end{figure}

\noindent Figure \ref{fig:4} presents the variation of flux function $-\Pi _{kv} $ with group size $n_{p} $ that is computed using simulation data at two different redshifts using dark matter flow dataset \citep{Xu:2022-Dark_matter-flow-dataset-part1,Xu:2022-Dark_matter-flow-dataset-part2}. The negative value of  $\Pi _{kv} $ indicates an inverse cascade of virial kinetic energy. There is also a scale-independent flux $\varepsilon _{kv} $ for small halos in mass propagation range. The virial kinetic energy decreases with time for a given size of halo ($\sigma _{v}^{2} $ decreases with time in Eq. \eqref{ZEqnNum222387} with $s=-1$), while the halo kinetic energy increases with time with $s=1$ for $\sigma _{h}^{2} $ in Eq. \eqref{ZEqnNum259105}. The flux function $\varepsilon _{kv} \approx {5/\left(3-2\tau _{0} \right)} \varepsilon _{kh} $ (Eqs. \eqref{ZEqnNum894335} and \eqref{ZEqnNum316450}), while the total amount of two kinetic energies in all halos are comparable with $\left\langle \sigma _{v}^{2} \right\rangle \approx \left\langle \sigma _{h}^{2} \right\rangle $ (Fig. \ref{fig:10}) at the statistically steady state.  

\subsection{Direct cascade of halo potential energy}
\label{sec:4.4}
The potential energy of halos can be similarly separated into two contributions: i) the potential  $\phi _{h}^{} \left(m_{h} ,a\right)$ due to inter-halo interactions between particles from different halos, and ii) the potential $\phi _{v}^{} \left(m_{h} ,a\right)$ due to intra-halo interactions between particles from the same halo. To better understand two different potentials, let's write the potential energy of a single particle that has two separate contributions (see decomposition of velocity dispersion in Eq. \eqref{ZEqnNum425919}), 
\begin{equation} 
\label{ZEqnNum622328} 
m_{p} \phi _{i} =\frac{1}{2} \sum _{j\ne i}^{N}V\left(r_{ji} \right) =\phi _{iv} +\phi _{ih} ,         
\end{equation} 
and
\begin{equation}
\phi _{iv} =\frac{1}{2} \sum _{k\ne i}^{n_{p} }V\left(r_{ki} \right) \quad \textrm{and} \quad \phi _{ih} =\frac{1}{2} \sum _{l}^{N-n_{p} }V\left(r_{li} \right),      
\label{ZEqnNum907599}
\end{equation}

\noindent respectively, where \textit{N} is the total number of particles in the system, $\phi _{i} $ is the specific potential of particle \textit{i}, $r_{ji} $ is the distance between two particles \textit{j} and \textit{i,} and $V\left(r_{ji} \right)$ is the gravitational potential. The first term $\phi _{iv} $ is the intra-halo potential from pair interactions with all other particles in the same halo of size $n_{p} $. 

The second term $\phi _{ih} $ is the inter-halo potential from pair interactions with all other particles (denoted as particle \textit{l}) out of the halo that particle \textit{i} resides in. Halos are small when compared to the typical separation between halos. Any two particles \textit{i} and \textit{j} in the same halo should have a distance $r_{ij} \ll r_{li} \approx r_{lj} $, where intra-halo distance is much smaller than the inter-halo distance. Therefore, particles in the same halo should have a similar background potential ($\phi _{ih} \approx \phi _{jh} $). 

The inter- and intra- halo potentials for individual halo are defined as the average of particle potentials for all $n_{p} $ particles in same halo, 
\begin{equation}
\begin{split}
&\phi _{hv} =\left\langle \phi _{iv} \right\rangle _{h} =\frac{1}{n_{p} } \sum _{i=1}^{n_{p} }\phi _{iv}\\ 
&\textrm{and}\\
&\phi _{hh} =\left\langle \phi _{ih} \right\rangle _{h} =\frac{1}{n_{p} } \sum _{i=1}^{n_{p} }\phi _{ih}  \approx \phi _{ih}.   
\label{eq:51}
\end{split}
\end{equation}

\noindent Finally, the inter- and intra- halo potentials for halo groups are defined as the average of all $n_{h} $ halos in the same group, i.e.
\begin{equation}
\begin{split}
&\phi _{v} \left(m_{h} ,a\right)=\left\langle \phi _{hv} \right\rangle _{g} =\frac{1}{n_{h} } \sum _{i=1}^{n_{h} }\phi _{hv}\\
&\textrm{and}\\ 
&\phi _{h} \left(m_{h} ,a\right)=\left\langle \phi _{hh} \right\rangle _{g} =\frac{1}{n_{h} } \sum _{i=1}^{n_{h} }\phi _{hh}.
\end{split}
\label{eq:52}
\end{equation}

\noindent Halo mass is small compared to the system mass ($m_{h} \ll M$ or $n_{p} \ll N$), such that the number of particles (particle \textit{l} in Eq. \eqref{ZEqnNum907599}) out of any halo is about the same for all different halos. Therefore, we expect the inter-halo potential $\phi _{h} $ for a halo group to be much less dependent on the halo size $m_{h} $ or $n_{p}$. By contrast, the intra-halo potential $\phi _{v}$ is strongly dependent on halo size $m_{h}$. At large scale, halos can be considered as macro-particles with a relatively homogeneous distribution in space. The inter-halo potential $\phi _{h}$ should be independent on both the position and mass of halos. Hence, $\phi _{h}$ may be considered as a background potential same for all halos. 

\begin{figure}
\includegraphics*[width=\columnwidth]{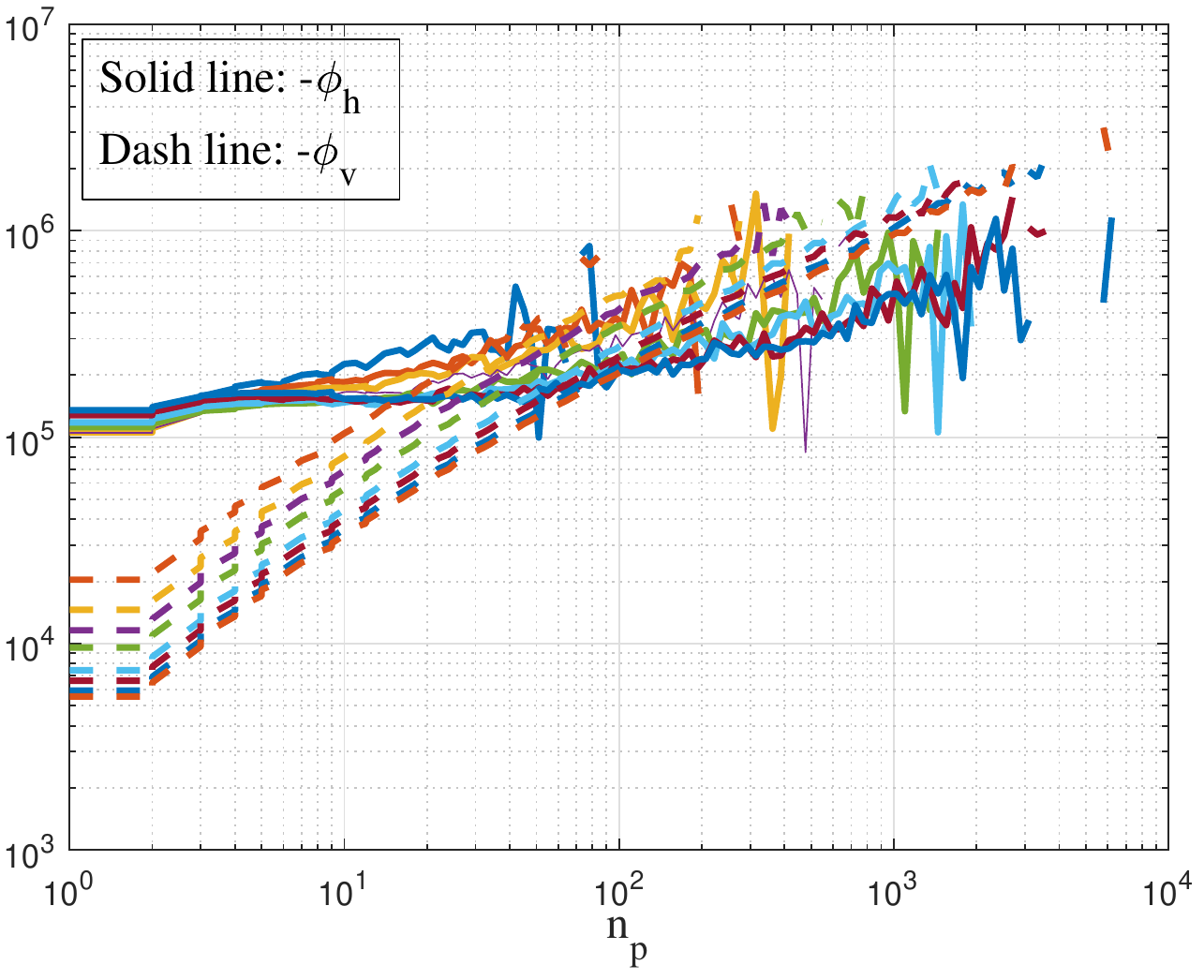}
\caption{The variation of inter-halo potential $\phi _{h} \left(m_{h} ,a\right)$ and intra-halo potential $\phi _{v} \left(m_{h} ,a\right)$ ($(Km/s)^2$) with halo group size $n_{p} $ for different redshifts \textit{z }= 0, 0.1, 0.3, 0.5, 1.0, 1.5, 2.0, and 3.0. The inter-halo potential $-\phi _{h} \sim a$ is relatively independent of the halo size, while inter-halo potential $\phi _{v} $ scales as $-\phi _{v} \sim a^{-1} m_{h}^{{2/3} } $ for large halos.}
\label{fig:5}
\end{figure}

Figure \ref{fig:5} plots the variation of intra-halo potential ($\phi _{v} $:dash lines) and inter-halo potential ($\phi _{h} $:solid lines) with halo group size $n_{p}$ at different redshifts \textit{z}. As expected, the inter-halo potential $-\phi _{h} \sim a$ is relatively independent of the halo size, while intra-halo potential $\phi _{v} $ scales as $-\phi _{v} \sim a^{-1} m_{h}^{{2/3}} $ for large halos. 

\begin{figure}
\includegraphics*[width=\columnwidth]{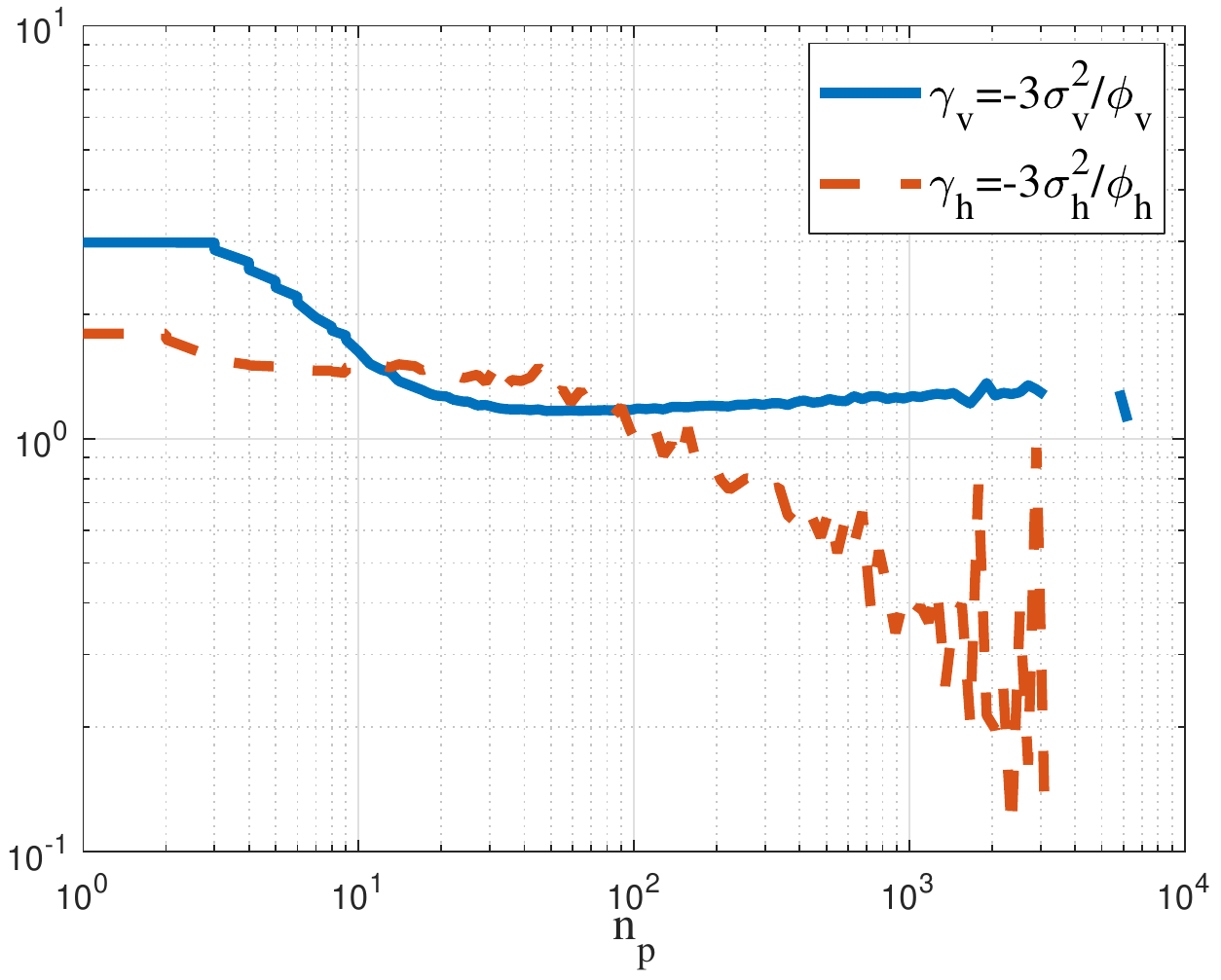}
\caption{The variation of two virial ratios $\gamma _{h} $ and $\gamma _{v} $ with halo group size $n_{p}$ for redshift z = 0. For random motion of halos, the ratio $\gamma _{h} $ approaches a constant value around 2 for $n_{p} =1$ (out-of-halo sub-system). For motion of particles in halos, the virial ratio $\gamma _{v} $ approaches a constant value between 1.3 and1.4 for large halos. Both ratios should be one if the virial theorem is exactly satisfied. The deviation of $\gamma _{v} $ is due to the halo surface energy (discussed in \citep{Xu:2021-Inverse-mass-cascade-halo-density}).}
\label{fig:6}
\end{figure}

The virial equilibrium for halo groups can be checked by the virial ratios defined as
\begin{equation}
\gamma _{v} =-{3\sigma _{v}^{2} / \phi _{v}^{} } \quad  \textrm{and} \quad \gamma _{h} =-{3\sigma _{h}^{2}/\phi _{h}^{} }     
\label{ZEqnNum611945}
\end{equation}

\noindent for motion in halos and motion of halos, respectively. Figure \ref{fig:6} plots the variation of two ratios $\gamma _{h} $ and $\gamma _{v} $ with halo group size $n_{p} $. For random motion of halos, $\gamma _{h} $ approaches a constant value 2 for $n_{p} =1$ (out-of-halo sub-system) and decreases with halo size. While for motion in halo, $\gamma _{v} $ approaches a constant value between 1.3 and 1.4 for large halos. This is due to the mass cascade, halo surface energy, and effective potential exponent $n_{e}$ \citep[see][Eq. (96)]{Xu:2021-Inverse-mass-cascade-halo-density}. There might be a greater uncertainty of $\gamma _{h} $ for large halos because of fewer number of large halos available for a reliable statistics.

Similar to Eq. \eqref{ZEqnNum593887}, the flux function for inter- and intra- potential energies are 
\begin{equation} 
\varepsilon _{\phi h} =\Pi _{\phi h} \left(0,a\right)=\varepsilon _{m} \left\langle \phi _{h} \right\rangle =-\left(\frac{3}{2} -\tau _{0} \right)M_{h} \left(a\right)H\left\langle \phi _{h} \right\rangle ,     
\label{ZEqnNum739926} 
\end{equation} 
\begin{equation} 
\varepsilon _{\phi v} =\Pi _{\phi v} \left(0,a\right)=\frac{\varepsilon _{m} \left\langle \phi _{v} \right\rangle }{{3/2} -\tau _{0} } \left[1+\frac{\partial \ln \left[-M_{h} \left(a\right)\left\langle \phi _{v} \right\rangle \right]}{\partial \ln a} \right].
\label{ZEqnNum452079} 
\end{equation} 

\begin{figure}
\includegraphics*[width=\columnwidth]{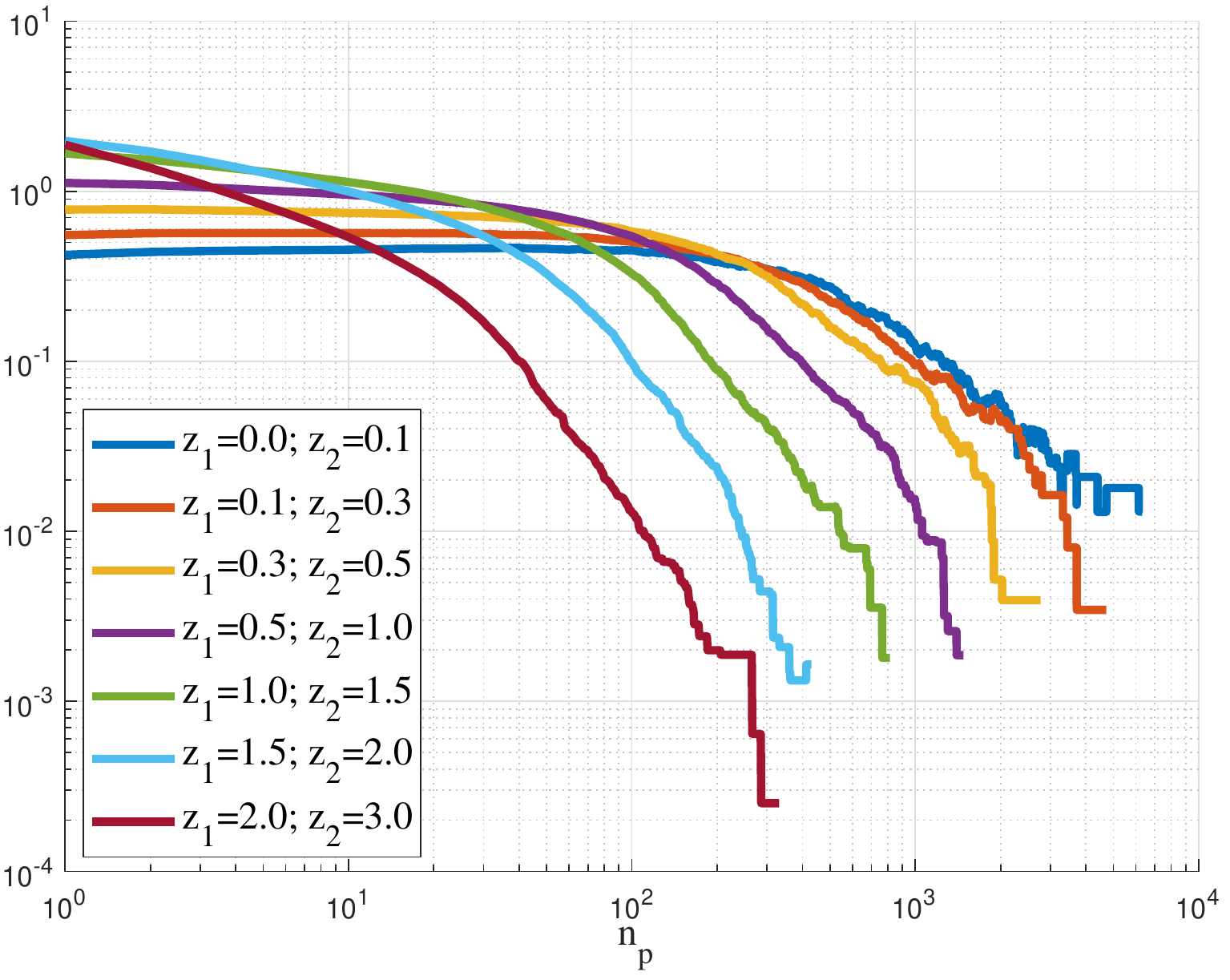}
\caption{The variation of flux function $\Pi _{\phi h} $ of the inter-halo potential energy $\phi _{h}^{} \left(m_{h} ,a\right)$ with size $n_{p} $ of halo groups. The flux function $\Pi _{\phi h} >0$ (direct cascade from large to small mass scales) is normalized by ${Nm_{p} u_{0}^{2}/t_{0} }$. A scale-independent flux function $\varepsilon _{\phi h} $ can be identified for mass propagation range with $m_{h} <m_{h}^{*} $.}
\label{fig:7}
\end{figure}

\begin{figure}
\includegraphics*[width=\columnwidth]{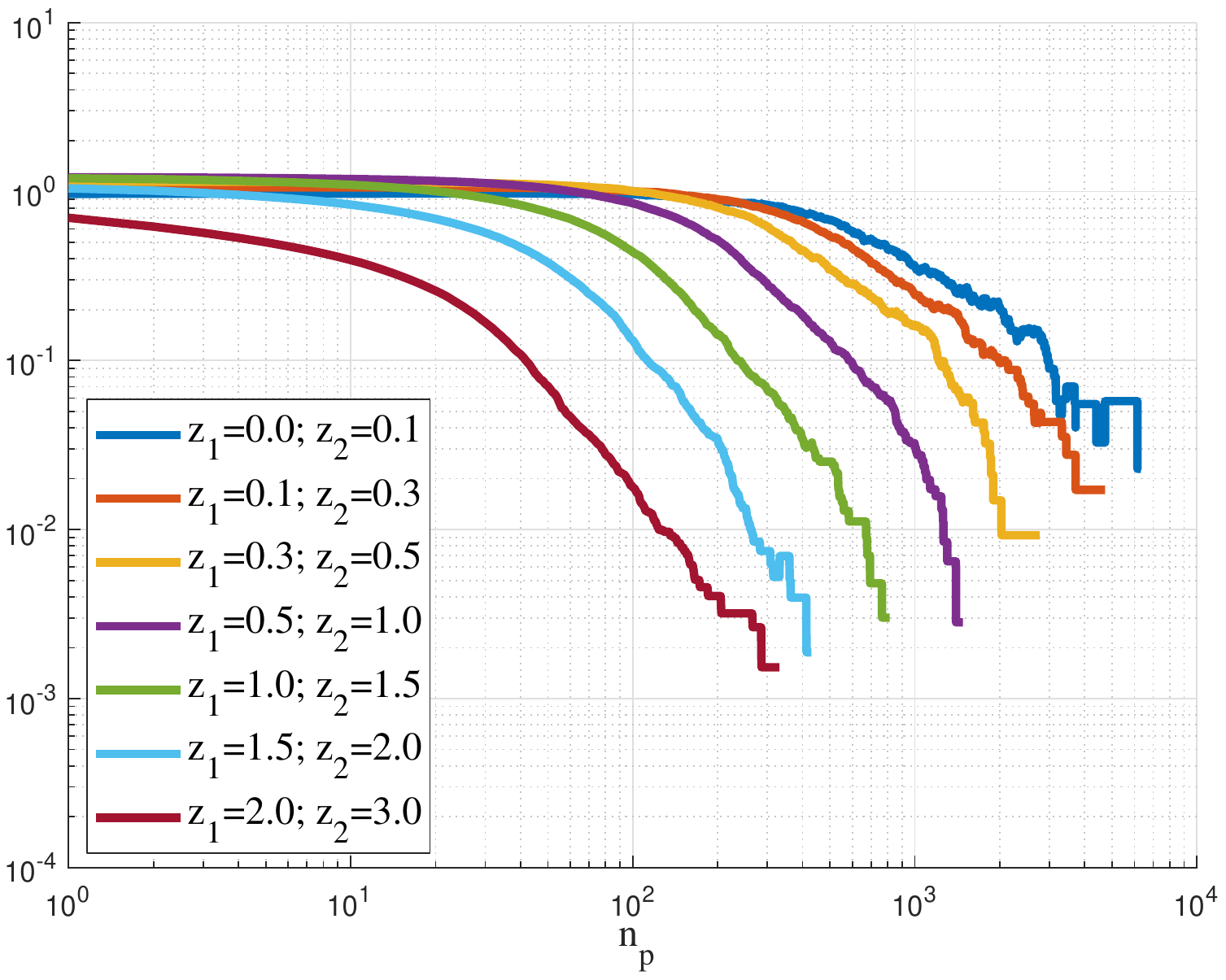}
\caption{The variation of flux function $\Pi _{\phi v} $ of the intra-halo potential energy $\phi _{v}^{} \left(m_{h} ,a\right)$ with size $n_{p} $ of halo groups. The flux function $\Pi _{\phi v} >0$ (direct cascade from large to small mass scales) is normalized by ${Nm_{p} u_{0}^{2}/t_{0} } $. A scale-independent flux function $\varepsilon _{\phi v} $ can be identified for mass propagation range with $m_{h} <m_{h}^{*}$.}
\label{fig:8}
\end{figure}

\noindent Figures \ref{fig:7} and \ref{fig:8} plot the variation of flux functions $\Pi _{\phi h} $ and $\Pi _{\phi v} $ for inter- and intra-halo potential energies, respectively. Both potential energies have a positive flux function indicating a direct cascade of potential energies, i.e. potential energy decreases (increases in absolute value) with halo growing from small to large mass scales. This can be shown from Eqs. \eqref{ZEqnNum739926} and \eqref{ZEqnNum452079}, where $\varepsilon _{m} <0$ and mean potential $\left\langle \phi _{h} \right\rangle <0$ and $\left\langle \phi _{v} \right\rangle <0$. Since both $\left\langle \phi _{h} \right\rangle $ and $\left\langle \phi _{v} \right\rangle $ can be dependent on $\left\langle \sigma _{h}^{2} \right\rangle $ and $\left\langle \sigma _{v}^{2} \right\rangle $ through virial theorem (Eq. \eqref{ZEqnNum611945}), the cascade of potential and kinetic energy are related to each other. The direct cascade of potential energy to the smallest mass scale provides the energy flux for the inverse cascade of kinetic energy.  

\subsection{The temporal evolution of kinetic and potential energies}
\label{sec:4.5}
In forced steady turbulence, the total mass is conserved, and total kinetic energy of entire system is also conserved. Kinetic energy is injected at the integral scale (the largest scale where external force is applied) and dissipated at the smallest scale (Kolmogorov scale) by molecular viscosity. However, the energy evolution in dark matter flow is much more complicated involving kinetic and potential energies, and virilization in halos. The total mass of all halos $M_{h} \left(a\right)$ also continuously increases due to the inverse mass cascade. 

Here we first divide entire system into a halo sub-system with a total mass of $M_{h} \left(a\right)$ and an out-of-halo sub-system with a total mass of $M_{o} \left(a\right)$. In simulation, halos are identified by a FoF algorithm with the smallest halo containing at least two particles. The total halo mass $M_{h} \left(a\right)$ includes mass of all particles from all halos. The total mass $M_{o} \left(a\right)$ of out-of-halo sub-system includes mass of all particles that does not belong to any halos. 

Second, the kinetic and potential energies for a halo sub-system can be decomposed on two different levels: i) halo kinetic energy $\left\langle \sigma _{h}^{2} \right\rangle $ and inter-halo potential $\left\langle \phi _{h}^{} \right\rangle $ for random motion of halos, and ii) virial kinetic energy $\left\langle \sigma _{v}^{2} \right\rangle $ and intra-halo potential $\left\langle \phi _{v}^{} \right\rangle $ for particle motion in halos. For out-of-halo sub-system, the mean kinetic energy $\left\langle \sigma _{ho}^{2} \right\rangle $ and potential energy $\left\langle \phi _{ho}^{} \right\rangle $ can also be computed with all out-of-halo particles identified. Here $\left\langle \right\rangle $ stands for the average of a quantity over all particles in respective sub-systems. 

\begin{figure}
\includegraphics*[width=\columnwidth]{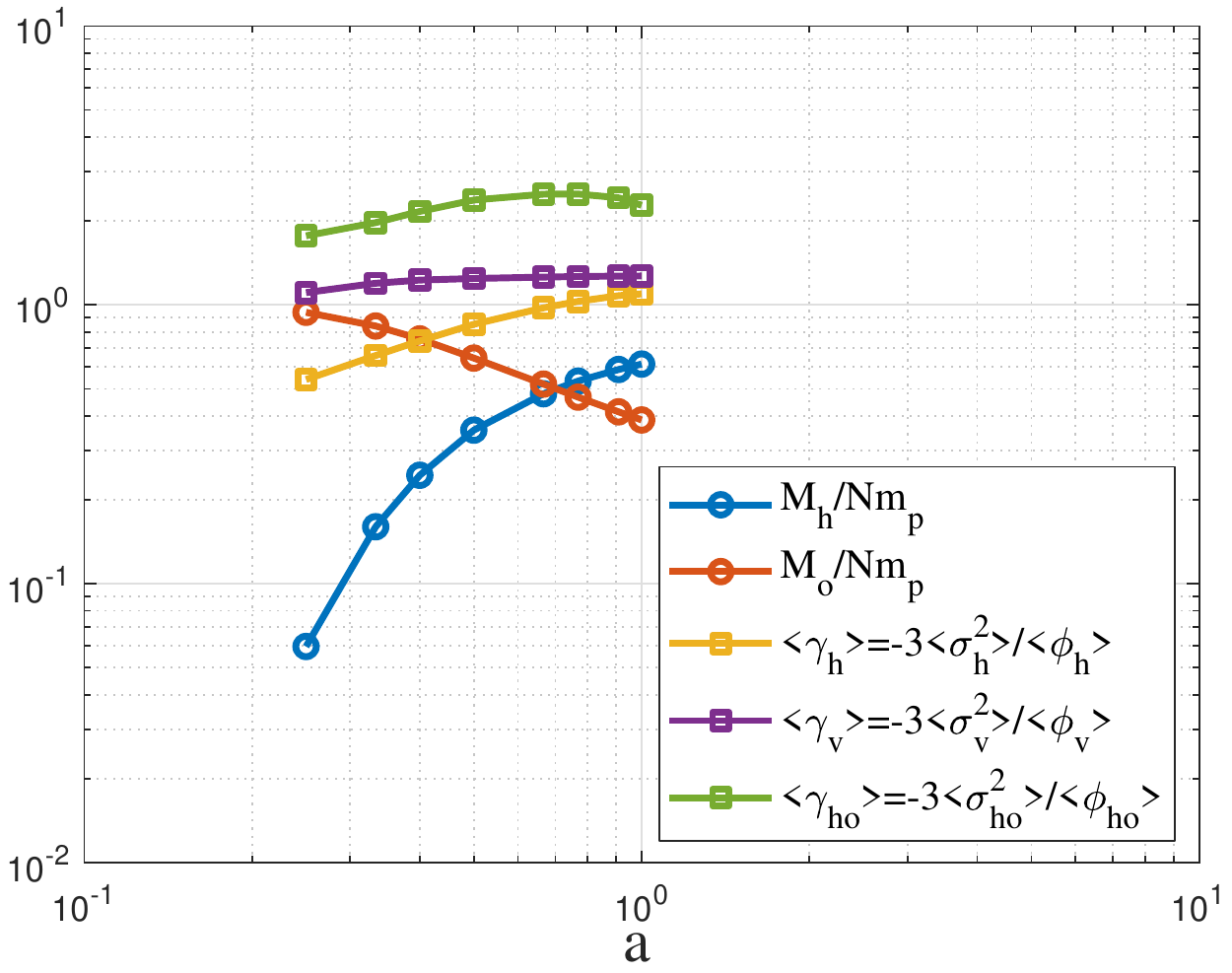}
\caption{The variation of halo mass $M_{h} \left(a\right)$ and out-of-halo mass $M_{O} \left(a\right)$ with scale factor \textit{a}. A continuous mass exchange between two sub-systems is required to sustain the total halo mass growing as $M_{h} \left(a\right)\sim a^{{1/2} } $ after reaching the statistically steady state. The variation of virial ratios for two sub-systems are also presented. For the motion of halos, the ratio $\left\langle \gamma _{h} \right\rangle $ (yellow) slowly increases from 0.5 to 1 and it takes longer time to reach virial equilibrium. The ratio $\left\langle \gamma _{v} \right\rangle \sim 1.3$ (purple) for particle motion within individual halos is relatively time-invariant and the virial equilibrium is established much faster in individual halos. The deviation of $\left\langle \gamma _{v} \right\rangle $ from 1 reflects the effects of halo mass accretion and halo surface energy \citep{Xu:2021-Inverse-mass-cascade-halo-density}. The virial ratio $\left\langle \gamma _{ho} \right\rangle \approx 2$ (green) for out-of-halo particles indicating that the out-of-halo sub-system is energy conserved (no virilization).}
\label{fig:9}
\end{figure}

Figure \ref{fig:9} plots the variation of $M_{h} \left(a\right)$ and $M_{o} \left(a\right)$ with scale factor \textit{a}. A continuous mass exchange ($M_{o} $ decreases while $M_{h} $ increases with \textit{a}) between two sub-systems is required to sustain the growth of total halo mass as $M_{h} \left(a\right)\sim a^{{1/2} } $ once the statistically steady state is established. 

The virial equilibrium for two sub-systems at two different levels (motion of halos and motion in halos) are also checked using virial ratios defined in Eq. \eqref{ZEqnNum611945}. At the halo level, $\left\langle \gamma _{h} \right\rangle $ slowly increases from $\mathrm{\sim}$0.5 to $\mathrm{\sim}$1 and the motion of halos takes longer time to reach virial equilibrium due to larger distance between halos and weaker gravity. For particle motion in halos, the virial ratio $\left\langle \gamma _{v} \right\rangle \sim 1.3$ is relatively time-invariant. The virial equilibrium is established much faster and earlier for motion in halos due to stronger gravity. The deviation of $\left\langle \gamma _{v} \right\rangle $ from 1 reflects the effects of halo mass accretion and halo surface energy \citep{Xu:2021-Inverse-mass-cascade-halo-density}. The virial ratio $\left\langle \gamma _{ho} \right\rangle \sim 2$ for an out-of-halo particles indicates that the out-of-halo sub-system is energy conserved (no virilization and kinetic energy always cancels potential energy) with ${3\left\langle \sigma _{ho}^{2} \right\rangle/2} +\left\langle \phi _{ho}^{} \right\rangle \approx 0$.  

\begin{figure}
\includegraphics*[width=\columnwidth]{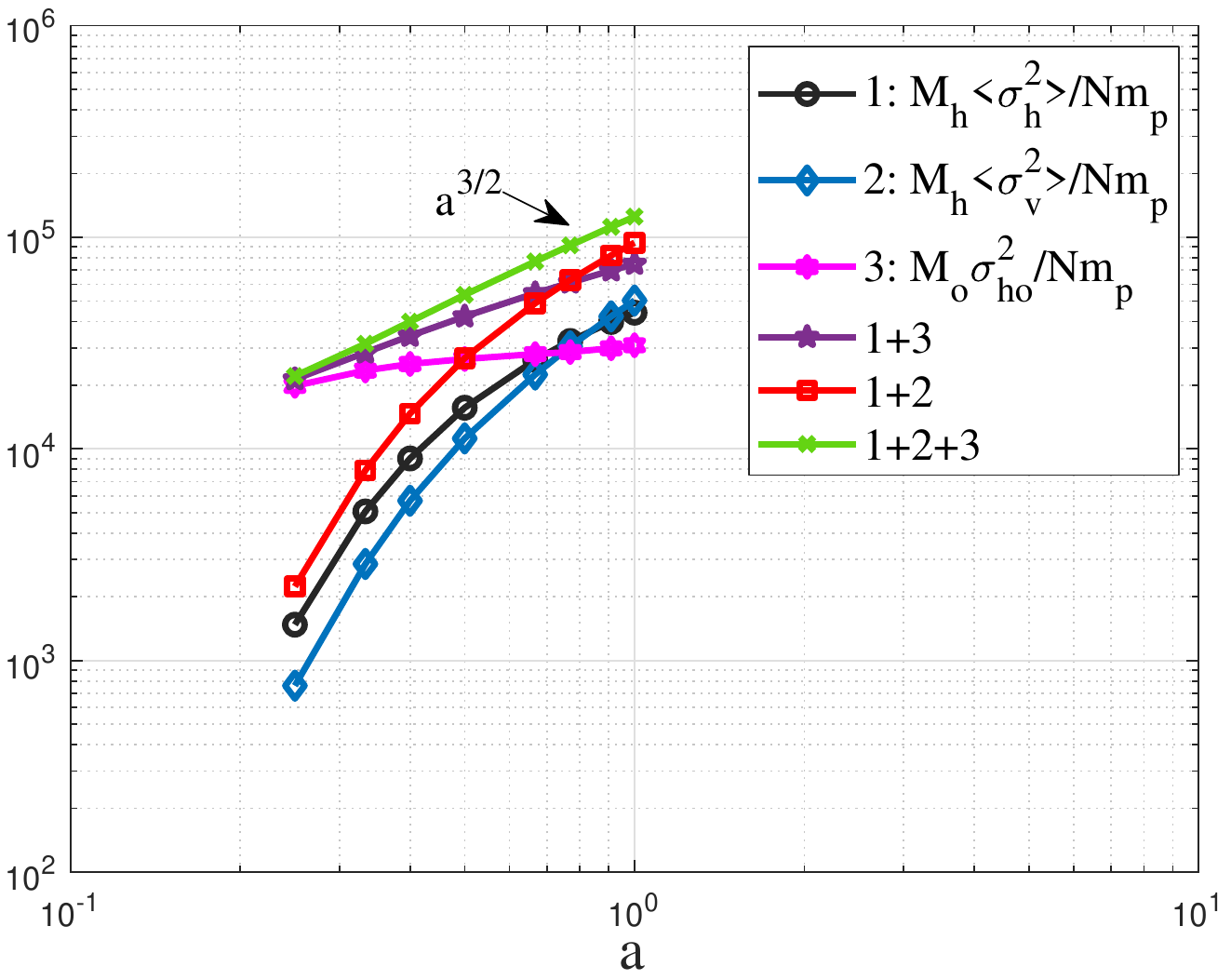}
\caption{The variation of three kinetic energies ($(Km/s)^2$) with scale factor \textit{a} i.e. halo kinetic energy (1: $\left\langle \sigma _{h}^{2} \right\rangle $ black), virial kinetic energy (2: $\left\langle \sigma _{v}^{2} \right\rangle $ blue), and out-of-halo kinetic energy (3: $\left\langle \sigma _{ho}^{2} \right\rangle $ magenta). The total kinetic energy of the entire system (green line: 1+2+3) grows approaching the scaling $a^{{3/2} } $. The kinetic energy of out-of-halo sub-system (magenta: 3) is relatively time-invariant. The total kinetic energy of halo sub-system (red: 1+2) becomes dominant over out-of-halo sub-system with $M_{h} \left(a\right)\sim a^{{1/2} } $, $\left\langle \sigma _{h}^{2} \right\rangle \sim a$, and $\left\langle \sigma _{v}^{2} \right\rangle \sim a$. A cross-over can be found at around \textit{a}=0.5.}
\label{fig:10}
\end{figure}

Figure \ref{fig:10} illustrates the variation of three kinetic energies with scale factor \textit{a}. The total kinetic energy of two sub-systems (green line: 1+2+3, i.e. the one-dimensional velocity dispersion $u^{2} $) grows approaching the scaling $\sim a^{{3/2}} $or $\mathrm{\sim}$\textit{t} (i.e. a constant energy flux function $\epsilon_u \sim u^2/t$, also see \citep{Xu:2022-The-evolution-of-energy--momen}). The total kinetic energy of out-of-halo sub-system (magenta: 3) is relatively time-invariant and does not changing with time. The total kinetic energy of out-of-halo sub-system is conserved while total mass $M_{o} $ decreases with time. The kinetic energy of halo sub-system (red: 1+2) becomes dominant over the out-of-halo sub-system with $M_{h} \left(a\right)\sim a^{{1/2} } $, $\left\langle \sigma _{h}^{2} \right\rangle \sim a$, and $\left\langle \sigma _{v}^{2} \right\rangle \sim a$. A cross-over can be found at around \textit{a}=0.5. 

\begin{figure}
\includegraphics*[width=\columnwidth]{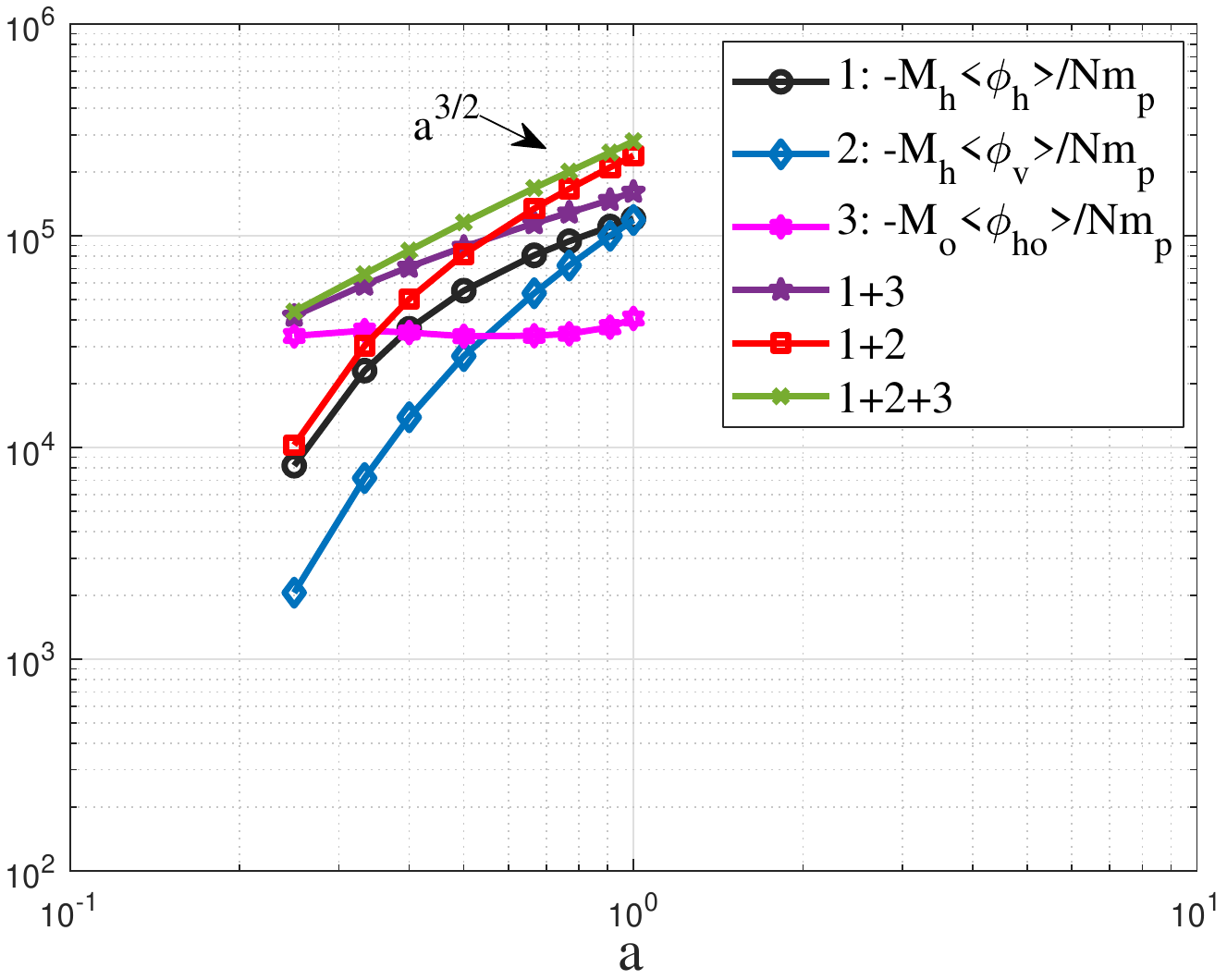}
\caption{The variation of three potential energies ($(Km/s)^2$) with scale factor \textit{a}, i.e. inter-halo potential (1: $\left\langle \phi _{h} \right\rangle $ black), intra-halo potential (2: $\left\langle \phi _{v} \right\rangle $ blue), and potential of out-of-halo particles (3: $\left\langle \phi _{ho} \right\rangle $ magenta). The total potential energy of two sub-systems (green line: 1+2+3) grows as $a^{{3/2} }$. The out-of-halo sub-system has a relatively time-invariant potential energy (magenta: 3). The potential energy from halo sub-system (red: 1+2) becomes dominant with increasing halo mass $M_{h} \left(a\right)\sim a^{{1/2} }$.}
\label{fig:11}
\end{figure}

Similarly, Fig. \ref{fig:11} shows the variation of three potential energies with scale factor \textit{a}. The total potential energy of two sub-systems (green line: 1+2+3) also grows as $a^{{3/2} } $. The out-of-halo sub-system has time-invariant potential energy (magenta: 3). The potential energy from halo sub-system (red: 1+2) becomes dominant after \textit{a}=0.5.

To summarize, there is a net mass flux into the halo sub-system to sustain the continuous growth of halo mass. However, the out-of-halo sub-system is energy conserved at any time, i.e. the total energy of out-of-halo sub-system (sum of potential and kinetic energy) is always nearly zero with a virial ratio $\gamma _{h} \approx 2$. The energy change of the entire system purely comes from the virilization in halo sub-system. The kinetic and potential energies of single mergers (particles in out-of-halo system) always cancel out. The inverse cascade of mass is accompanied by an inverse cascade of kinetic energy and a direct cascade of potential energy. Two energy cascades are related by the virial ratios. Total kinetic and potential energies (absolute value) of halo sub-system increase linearly with time \textit{t} when statistically steady state is established. This leads to a constant energy flux in halo sub-system.   

\subsection{Inverse cascade of halo radial and rotational kinetic energy}
\label{sec:4.6}
In hydrodynamic turbulence, the production of turbulence kinetic energy is facilitated through the Reynolds stress, a fictitious stress arising from velocity fluctuations to account for the effect of turbulence on mean flow. The Reynolds stress acts as a bridge for transferring kinetic energy from mean flow at large scales to turbulence and cascading down to the smallest scale. In this picture, the random motion (turbulence) is continuously drawing energy from the coherent motion (mean flow) at large scale. 

For dark matter flow (SG-CFD), there does not exist a mean flow on the largest scale, where it is assumed to be isotropic and homogeneous. on the scale of individual halos, the coherent motion (mean flow) of dark matter particles includes the halo radial and rotational motion. The kinetic energy associated with the coherent radial and rotational motion can be part of the virial kinetic energy $\sigma _{v}^{2} $ and similarly cascaded across halos of different mass scales, which will be studied in this section. A complete analysis of the coherent motion in halos and energy transfer with random motion requires full knowledge of mean flow and dispersion of halos. This is presented in a separate paper \citep{Xu:2022-The-mean-flow--velocity-disper}.

\begin{figure}
\includegraphics*[width=\columnwidth]{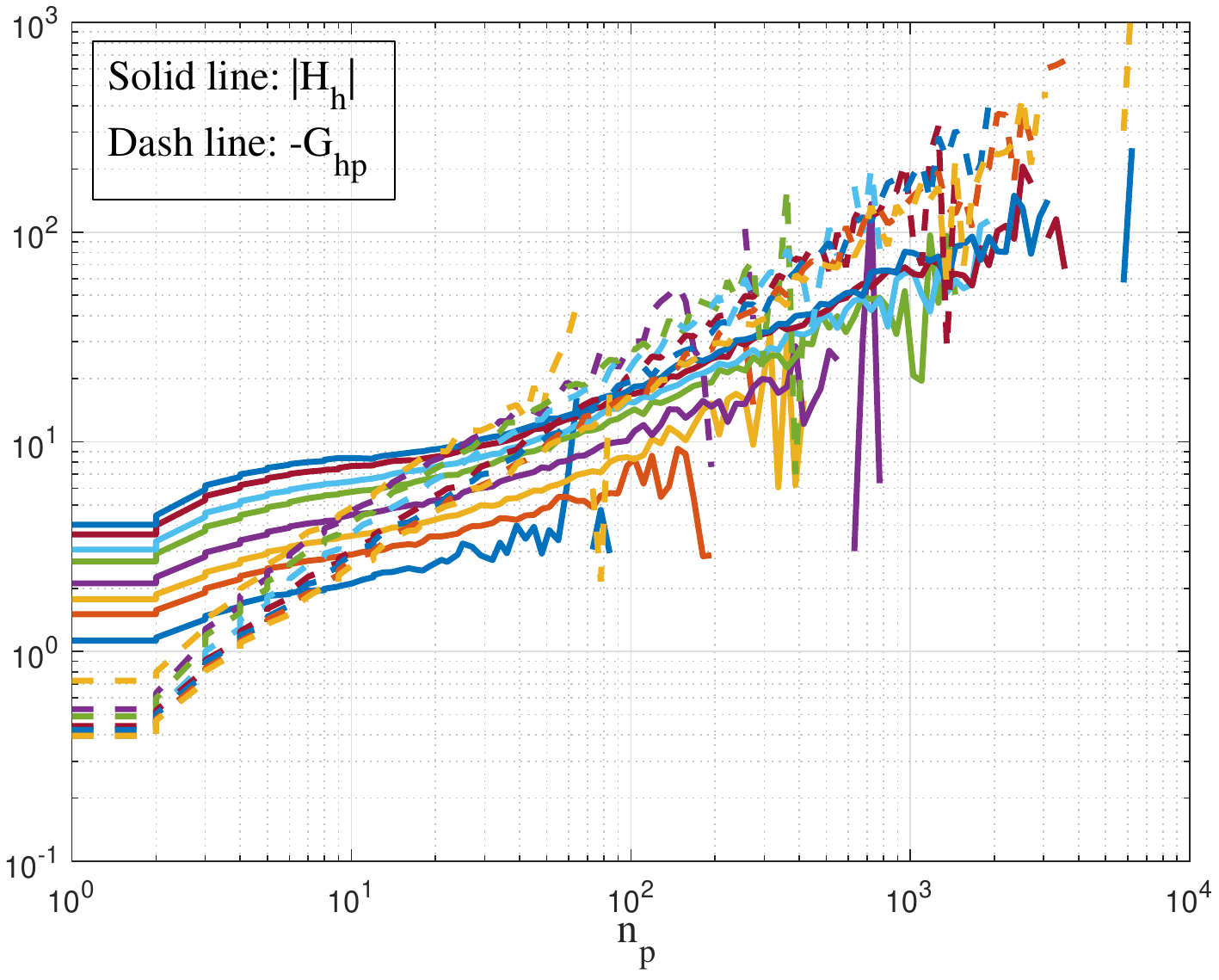}
\caption{The variation of halo peculiar virial quantity $G_{hp}$ and angular momentum $\left|\boldsymbol{\mathrm{H}}_{h}\right|$ ($Km/s \cdot Mpc/h$) with halo group size $n_{p}^{} $ for different redshifts \textit{z }= 0, 0.1, 0.3, 0.5, 1.0, 1.5, 2.0, and 3.0. For a give halo size, the virial quantity decreases with scale factor \textit{a} as $G_{hp} \approx -f_{G} \left(m_{h} \right)a^{-1} Hr_{g}^{2} $, while angular momentum increases with \textit{a} as $\left|\boldsymbol{\mathrm{H}}_{h} \right|\approx f_{H} \left(m_{h} \right)a^{{1/2} } Hr_{g}^{2}$.}
\label{fig:12}
\end{figure}

Just like the velocity vector (Eq. \eqref{ZEqnNum502045}), the particle position vector can be decomposed as $\boldsymbol{\mathrm{x}}_{p} =\boldsymbol{\mathrm{x}}_{h} +\boldsymbol{\mathrm{x}}_{p}^{'} $, where $\boldsymbol{\mathrm{x}}_{h} $ is the (physical) position vector of the center of halo mass that particle reside in and $\boldsymbol{\mathrm{x}}_{p}^{'} $ is the relative position of particles to center of mass. The halo radial and rotational motions can be defined by a halo (peculiar) virial quantity $G_{hp}$ (or radial momentum) and angular momentum $\boldsymbol{\mathrm{H}}_{h} $ in physical coordinate (not comoving coordinate), 
\begin{equation}
G_{hp} =\frac{1}{n_{p} } \sum _{i=1}^{n_{p} }\left(\boldsymbol{\mathrm{x}}_{p}^{'} \cdot \boldsymbol{\mathrm{u}}_{p}^{'} \right) \quad \textrm{and} \quad \boldsymbol{\mathrm{H}}_{h} =\frac{1}{n_{p} } \sum _{i=1}^{n_{p} }\left(\boldsymbol{\mathrm{x}}_{p}^{'} \times \boldsymbol{\mathrm{u}}_{p}^{'} \right),  
\label{eq:56}
\end{equation}

\noindent where $\boldsymbol{\mathrm{u}}_{p}^{'} $ is the peculiar velocity. Both quantities are first order moment of velocity. Figure \ref{fig:12} presents the variation of the peculiar halo virial quantity $G_{hp} $ and angular momentum $\left|\boldsymbol{\mathrm{H}}_{h} \right|$ with halo group size $n_{p}$, where both increase with the halo size. 

\begin{figure}
\includegraphics*[width=\columnwidth]{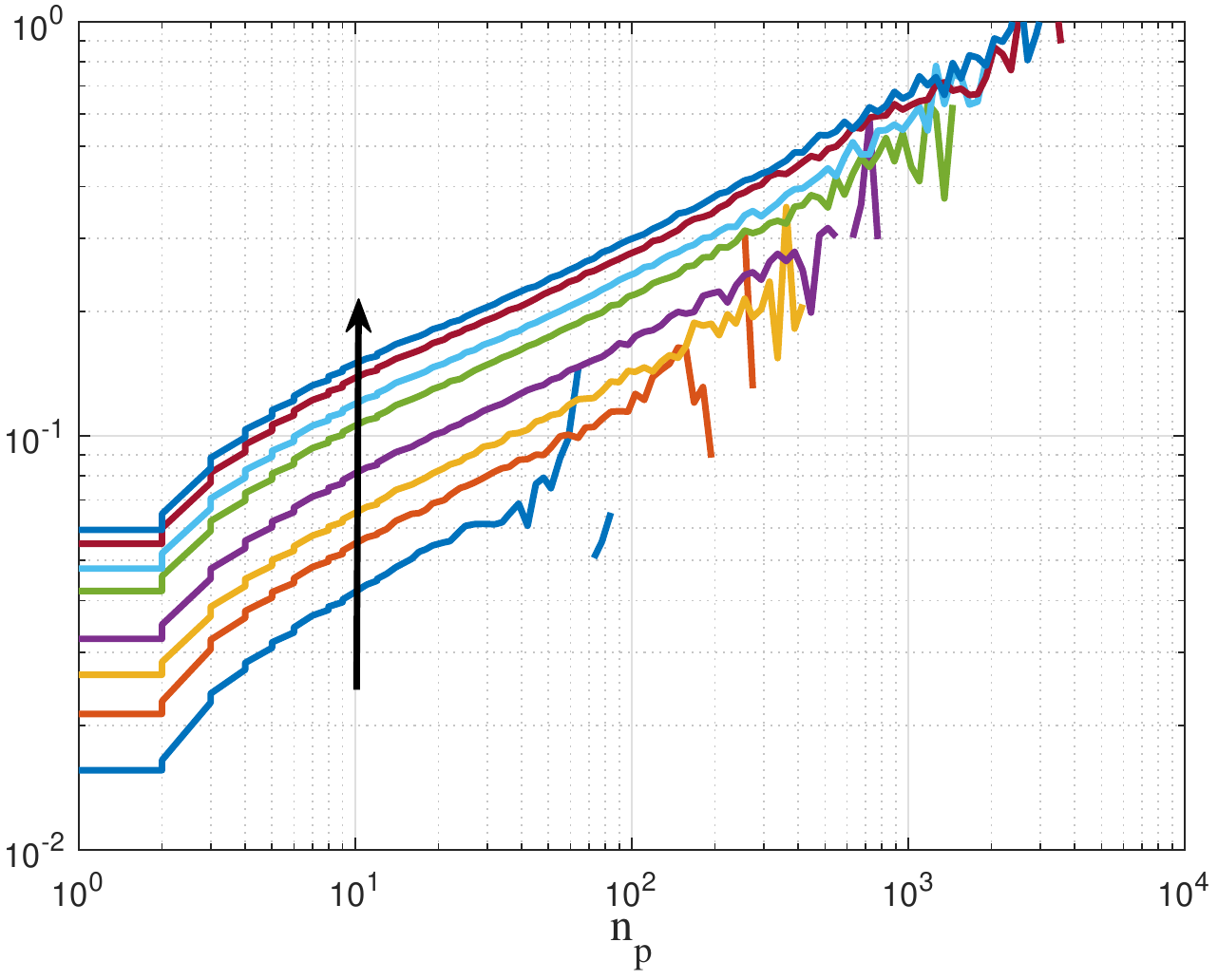}
\caption{The variation of (physical) mean square radius $r_{g}^{} \left(m_{h} ,a\right)$ (Mpc/h) with group size $n_{p}^{} $ for different redshifts \textit{z }= 0, 0.1, 0.3, 0.5, 1.0, 1.5, 2.0, and 3.0 with arrow pointing to z=0. The mean square radius $r_{g}^{} \sim am_{h}^{{1/3} } $ is proportional to \textit{a} and increasing with $m_{h} $.}
\label{fig:13}
\end{figure}

\begin{figure}
\includegraphics*[width=\columnwidth]{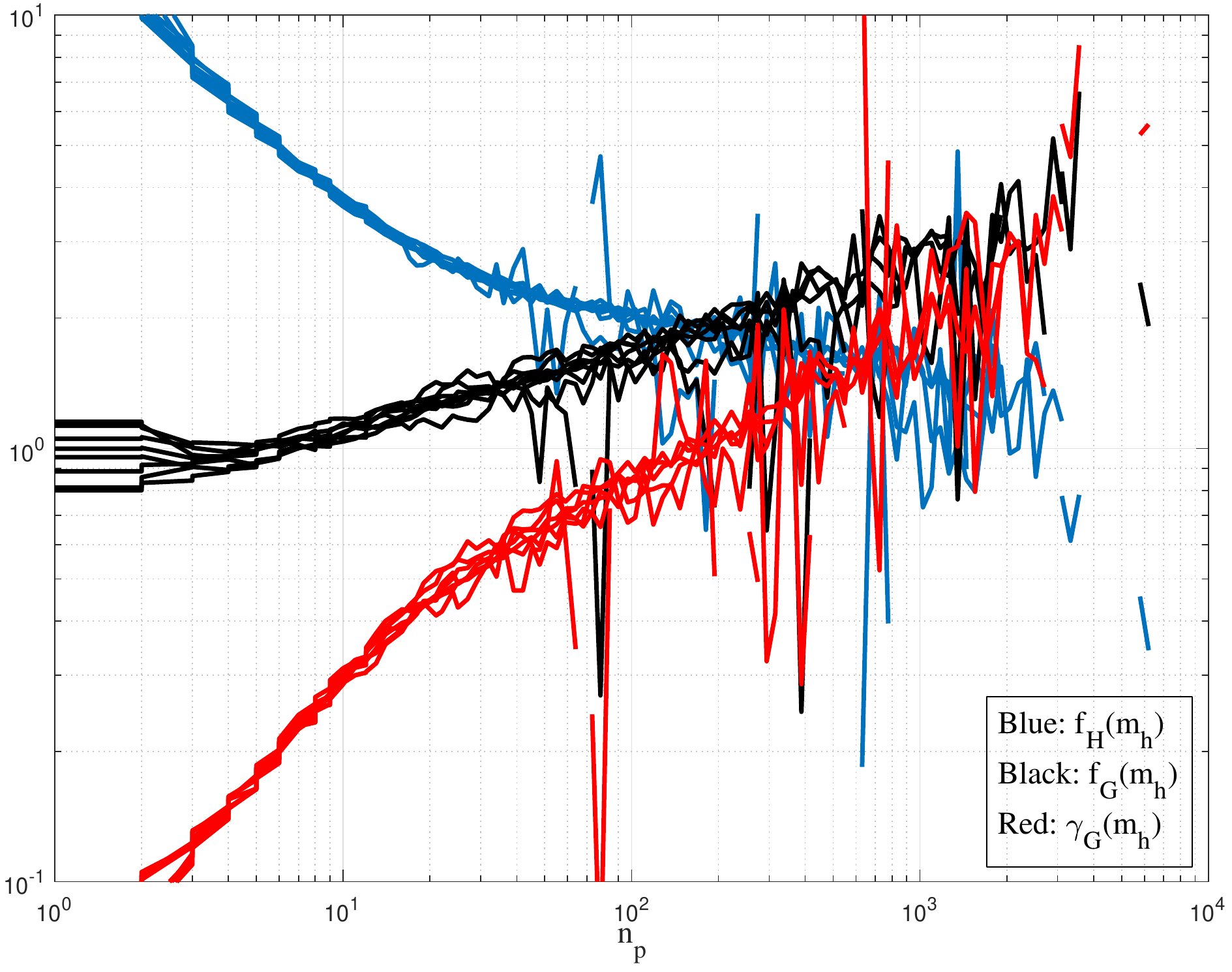}
\caption{The variation of ratios $\gamma _{G} $ (red), $f_{G} $ (black) and $f_{H} $ (blue) with halo group size $n_{p}$ for different redshifts \textit{z }= 0, 0.1, 0.3, 0.5, 1.0, 1.5, 2.0, and 3.0. All data from different redshifts collapse on a single line (no redshift dependence for these ratios). For small halos, $\left|\boldsymbol{\mathrm{H}}_{h}^{} \left(m_{h} ,a\right)\right|\propto -G_{hp}^{} \left(m_{h} ,a\right)a^{{3/2} } $. Note that $\gamma _{G} \approx {1/\left(3\pi \right)} $ at the smallest scale as predicted by a two-body collapse model (TBCM) \citep{Xu:2021-A-non-radial-two-body-collapse}. The ratio $\gamma _{G} $ and $f_{G} $ increases with halo size, while $f_{H}$ decreases with halo size.}
\label{fig:14}
\end{figure}

N-body simulations suggest the best expressions of, 
\begin{equation}
G_{hp} \approx -f_{G} \left(m_{h} \right)a^{-0.85} Hr_{g}^{2}, \quad \left|\boldsymbol{\mathrm{H}}_{h} \right|\approx f_{H} \left(m_{h} \right)a^{0.65} Hr_{g}^{2},  
\label{ZEqnNum249712}
\end{equation}

\noindent where $f_{G} \left(m_{h} \right)\in \left[1,3\right]$ and $f_{H} \left(m_{h} \right)\in \left[10,1\right]$ are two functions of halo mass $m_{h} $ only. It is expected that $f_{G} \left(m_{h} \right)$ increases with $m_{h} $ while $f_{H} \left(m_{h} \right)$ decreases with $m_{h} $ (Fig. \ref{fig:14}). The limiting values of $f_{G} \left(m_{h} \right)$ and $f_{H} \left(m_{h} \right)$ can be derived for large and small halos \citep[see][Table 3]{Xu:2022-The-evolution-of-energy--momen,Xu:2022-The-mean-flow--velocity-disper}. Another option for two quantities can be
\begin{equation}
G_{hp} \approx -f_{G} \left(m_{h} \right)a^{-1} Hr_{g}^{2}, \quad \left|\boldsymbol{\mathrm{H}}_{h} \right|\approx f_{H} \left(m_{h} \right)a^{0.5} Hr_{g}^{2},  
\label{ZEqnNum671144} 
\end{equation}

\noindent where the mean square radius $r_{g} $ for a given halo is defined as
\begin{equation} 
\label{eq:59} 
r_{g} =\sqrt{{\sum _{p=1}^{n_{p} }\left|\boldsymbol{\mathrm{x}}_{p}^{'} \right|^{2}/n_{p} } } .          
\end{equation} 
Here $\left|\boldsymbol{\mathrm{x}}_{p}^{'} \right|$ is the distance of the \textit{p}th particle to halo center of mass. Compared to halo size $r_{h} $ (virial radius that depends on the critical density ratio $\Delta _{c} $), the mean square radius $r_{g} $ is a well-defined quantity and easy to compute for every halo identified. For spherical halos of size $r_{h} $ with a power-law density of $\rho _{h} \left(r\right)\sim r^{-m} $, the mean square radius can be found as
\begin{equation} 
\label{ZEqnNum229152} 
r_{g} =\sqrt{\frac{\int _{0}^{r_{h} }r^{2} \rho _{h} \left(r\right)4\pi r^{2} dr }{\int _{0}^{r_{h} }\rho _{h} \left(r\right)4\pi r^{2} dr } } =\sqrt{\frac{3-m}{5-m} } r_{h} .        
\end{equation} 
Figure \ref{fig:13} plots the variation of $r_{g} $ with halo group size $n_{p}^{} $, where the mean square radius scales as $r_{g}^{} \sim am_{h}^{{1/3} } $. If  $r_g$ can be related to the halo size as $r_{g} =\gamma _{g} r_{h} $, where $\gamma _{g} $ is a dimensionless constant on the order of ${1/\sqrt{3} } $ for a spherical halo (Eq. \eqref{ZEqnNum229152}) with \textit{m}=2 for an isothermal density profile), we can conveniently write a formula for $r_g$ and an expression fitted from simulation (average halo density is about $\Delta _{c} $ times of the background density, i.e. $\bar{\rho }_{h} =\Delta _{c} \bar{\rho }_{0} $) 
\begin{equation}
\begin{split}
&r_{g}^{} =\gamma _{g} a\left(\frac{2Gm_{h} }{\Delta _{c} H_{0}^{2} } \right)^{{1/3}},\\ 
&\frac{r_{g}^{} }{{Mpc/h}} \approx 0.28a\left(\frac{m_{h}}{{2.27\times 10^{13} M_{\odot }/h}} \right)^{{1/3}}.
\end{split}
\label{ZEqnNum489297}
\end{equation}

\noindent The relation between $G_{hp}^{} $ and $\left|\boldsymbol{\mathrm{H}}_{h}^{} \right|$ was predicted to scale as $a^{{-3/2} } $ from a two-body collapse model \citep[see TBCM][Eq. (104)]{Xu:2021-A-non-radial-two-body-collapse}. Here we can define a ratio $\gamma _{G} $ between two quantities,  
\begin{equation} 
\label{eq:62} 
\gamma _{G} =\frac{-G_{hp}^{} \left(m_{h} ,a\right)a^{{3/2} } }{\left|\boldsymbol{\mathrm{H}}_{h}^{} \left(m_{h} ,a\right)\right|} =\frac{f_{G} \left(m_{h} \right)}{f_{H} \left(m_{h} \right)} ,        
\end{equation} 
which quantifies the relative importance of radial motion to rotational motion. Figure \ref{fig:14} shows the variation of the ratio $\gamma _{G} $ (red), $f_{G} $ (black) and $f_{H} $ (blue) in Eq. \eqref{ZEqnNum249712} with halo group size $n_{p}^{} $ for different redshifts \textit{z}. All data from different redshifts collapse on to a single line (no redshift dependence). Note that $\gamma _{G} \approx {1/\left(3\pi \right)} $ for the smallest halos as predicted by the two-body collapse model \citep[see][Eq. (104)]{Xu:2021-A-non-radial-two-body-collapse}. The ratio $\gamma _{G} $ increases with halo size. Rotational motion is dominant over the radial motion for small halos and two motions are comparable for large halos. The large uncertainty for large halos might come from the lack of reliable statistics for large halos.

The radius of gyration $r_{rg} $ about any axis for a spherical halo with a power-law density simply reads 
\begin{equation} 
\label{eq:63} 
r_{rg} =r_{h} \sqrt{\frac{2\left(3-m\right)}{3\left(5-m\right)} } =r_{g} \sqrt{\frac{2}{3} } .         
\end{equation} 
An effective halo angular velocity $\omega _{h} $ can be defined (using Eq. \eqref{ZEqnNum671144}),
\begin{equation} 
\label{ZEqnNum316976} 
\omega _{h} =\frac{\left|\boldsymbol{\mathrm{H}}_{h} \right|}{r_{rg}^{2} } =\frac{3\left|\boldsymbol{\mathrm{H}}_{h} \right|}{2r_{g}^{2} } =\frac{3}{2} f_{H} \left(m_{h} \right)H_{0} a^{-1} .       
\end{equation} 

\begin{figure}
\includegraphics*[width=\columnwidth]{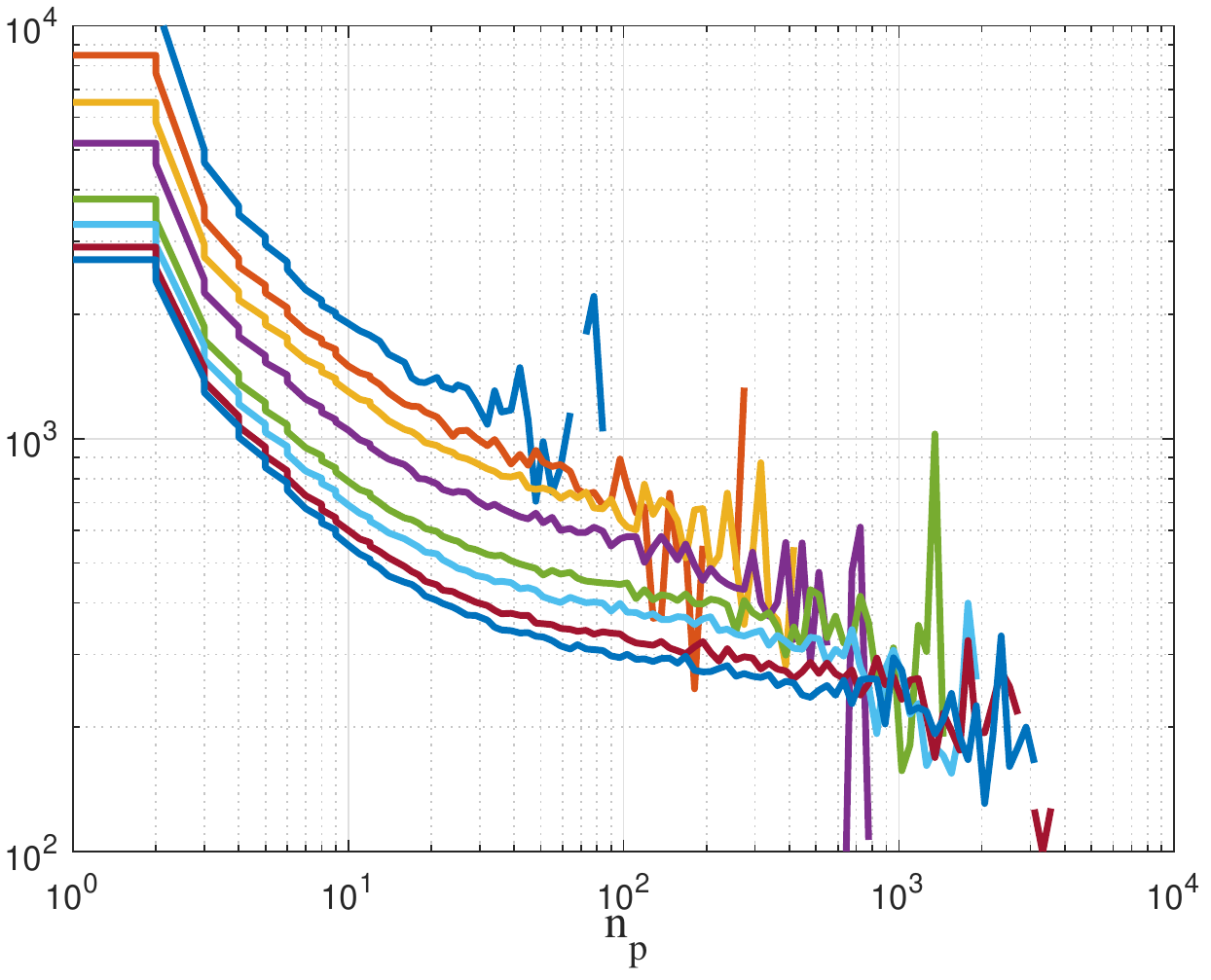}
\caption{The variation of effective angular velocity $\omega _{h}^{} \left(m_{h} ,a\right)$ ($(Km/s)/(Mpc/h)$) with halo size for different redshifts \textit{z }= 0, 0.1, 0.3, 0.5, 1.0, 1.5, 2.0, and 3.0. Small halos tend to rotate much faster, while the halo angular velocity approaches a constant value of $2\sim 3H$ for large halos.}
\label{fig:15}
\end{figure}
\noindent Figure \ref{fig:15} plots the variation of effective angular velocity $\omega _{h} $ with halo size $n_{p}^{} $. Small halos tend to have a greater angular velocity, while large halos approach a constant angular velocity between $2\sim 3H$. Exact values is presented for large halos \citep[see][Table 3]{Xu:2022-The-mean-flow--velocity-disper}. 

\begin{figure}
\includegraphics*[width=\columnwidth]{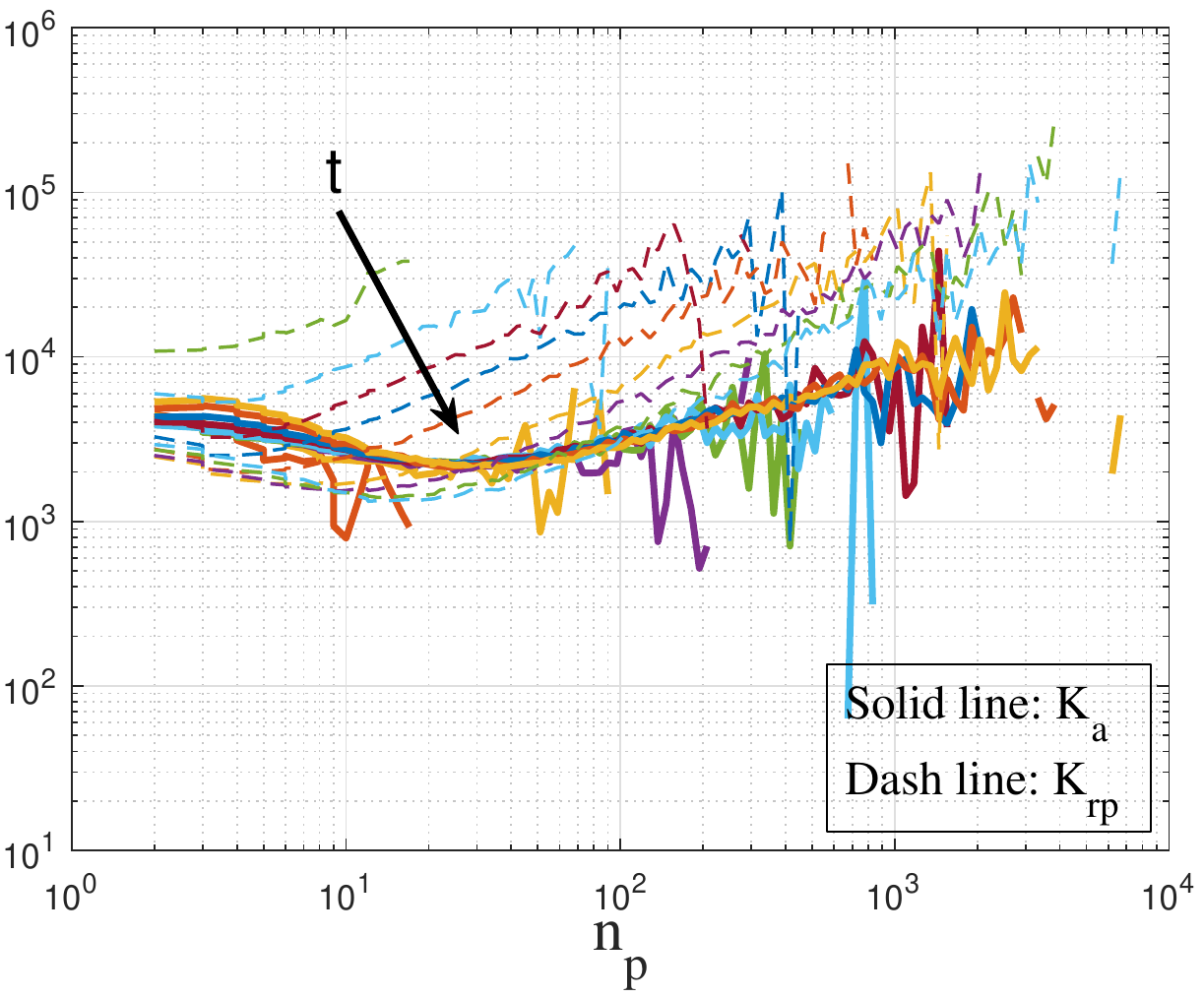}
\caption{The variation of radial ($K_{rp} $) and rotational kinetic energies ($K_{a} $) ($(Km/s)^2$) with halo group size $n_{p}^{} $ for different redshifts \textit{z }= 0, 0.1, 0.3, 0.5, 1.0, 1.5, 2.0, and 3.0. For a given size of halo, radial peculiar kinetic energy $K_{rp} $ is decreasing with scale factor \textit{a}, while rotational kinetic energy $K_{a} $ is relatively independent of time. Arrow points to \textit{z}=0.}
\label{fig:16}
\end{figure}

The (peculiar) radial and rotational kinetic energies can be approximated as (using Eqs. \eqref{ZEqnNum671144} and \eqref{ZEqnNum489297}),
\begin{equation} 
\label{ZEqnNum528733} 
K_{rp} =\frac{1}{2} \left(\frac{G_{hp}}{r_{g}} \right)^{2} =\frac{1}{2} \gamma _{g}^{2} a^{-3} \left[f_{G}\right]^{2} \left[\frac{2Gm_{h} H_{0} }{\Delta _{c} } \right]^{{2/3} } ,     
\end{equation} 
and 
\begin{equation} 
\label{ZEqnNum523492} 
K_{a} =\frac{1}{2} \left|\boldsymbol{\mathrm{H}}_{h} \right|\omega _{h} =\frac{3}{4} \left(\frac{\left|\boldsymbol{\mathrm{H}}_{h} \right|}{r_{g}} \right)^{2} =\frac{3}{4} \gamma _{g}^{2} \left[f_{H}\right]^{2} \left[\frac{2Gm_{h} H_{0} }{\Delta _{c} } \right]^{{2/3} } .    
\end{equation} 
Figure \ref{fig:16} presents the variation of the specific radial and rotational kinetic energies with halo size $n_{p}^{} $ for different redshifts \textit{z}. For a given size of halo, the radial kinetic energy $K_{rp} $ decreases with time, while the rotational kinetic energy $K_{a} $ is relatively independent of time. Both radial and rotational kinetic energy increase with halo size. 

\begin{figure}
\includegraphics*[width=\columnwidth]{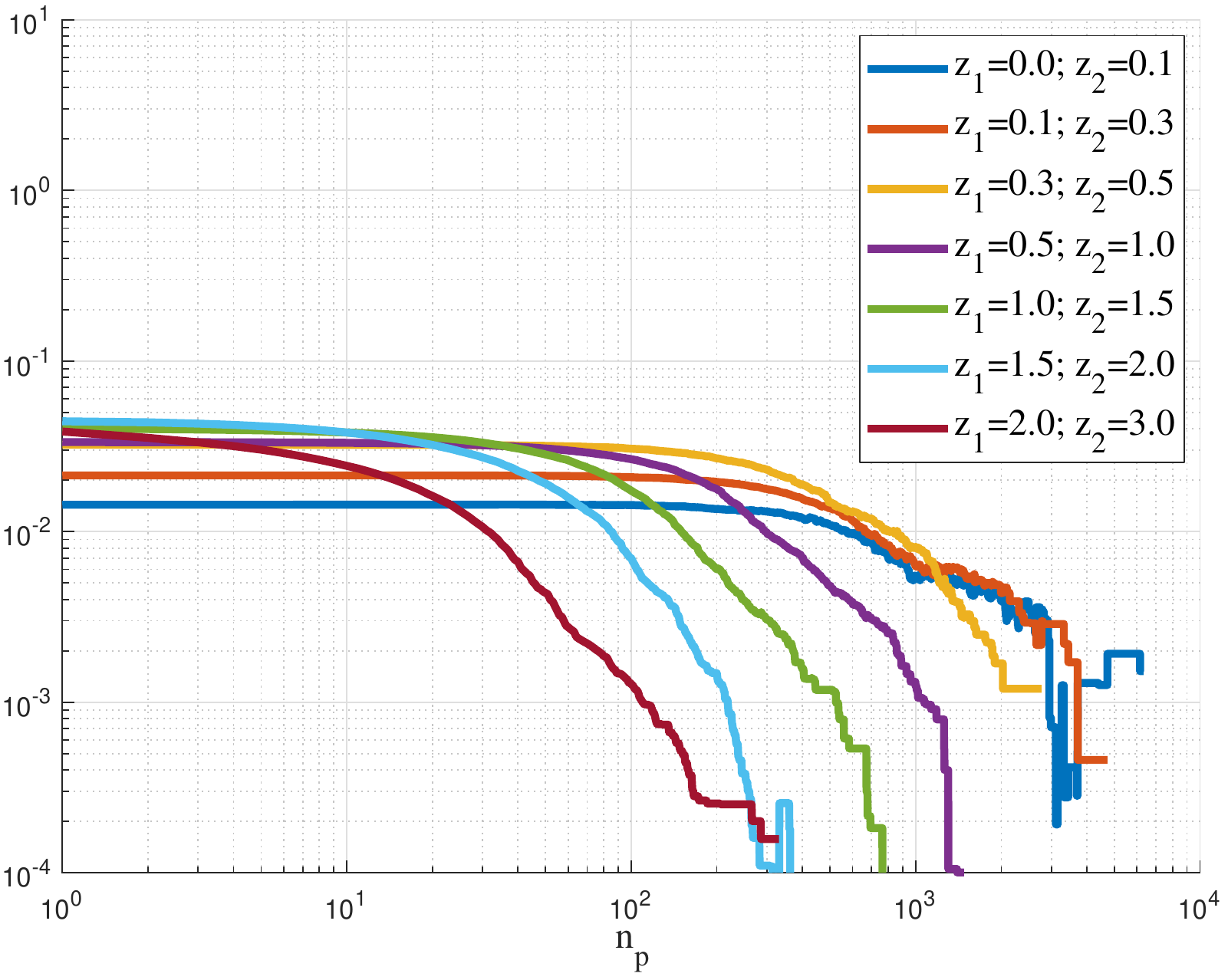}
\caption{The variation of flux function $\Pi _{ur} $ for halo radial kinetic energy $K_{rp}^{} \left(m_{h} ,a\right)$ with halo group size $n_{p} $. The flux function $\Pi _{ur} <0$ (inverse cascade from small to large scales) and is normalized by ${Nm_{p} u_{0}^{2}/t_{0} } $. A scale-independent flux function can be identified for mass propagation range with $m_{h} <m_{h}^{*} $.}
\label{fig:17}
\end{figure}

\begin{figure}
\includegraphics*[width=\columnwidth]{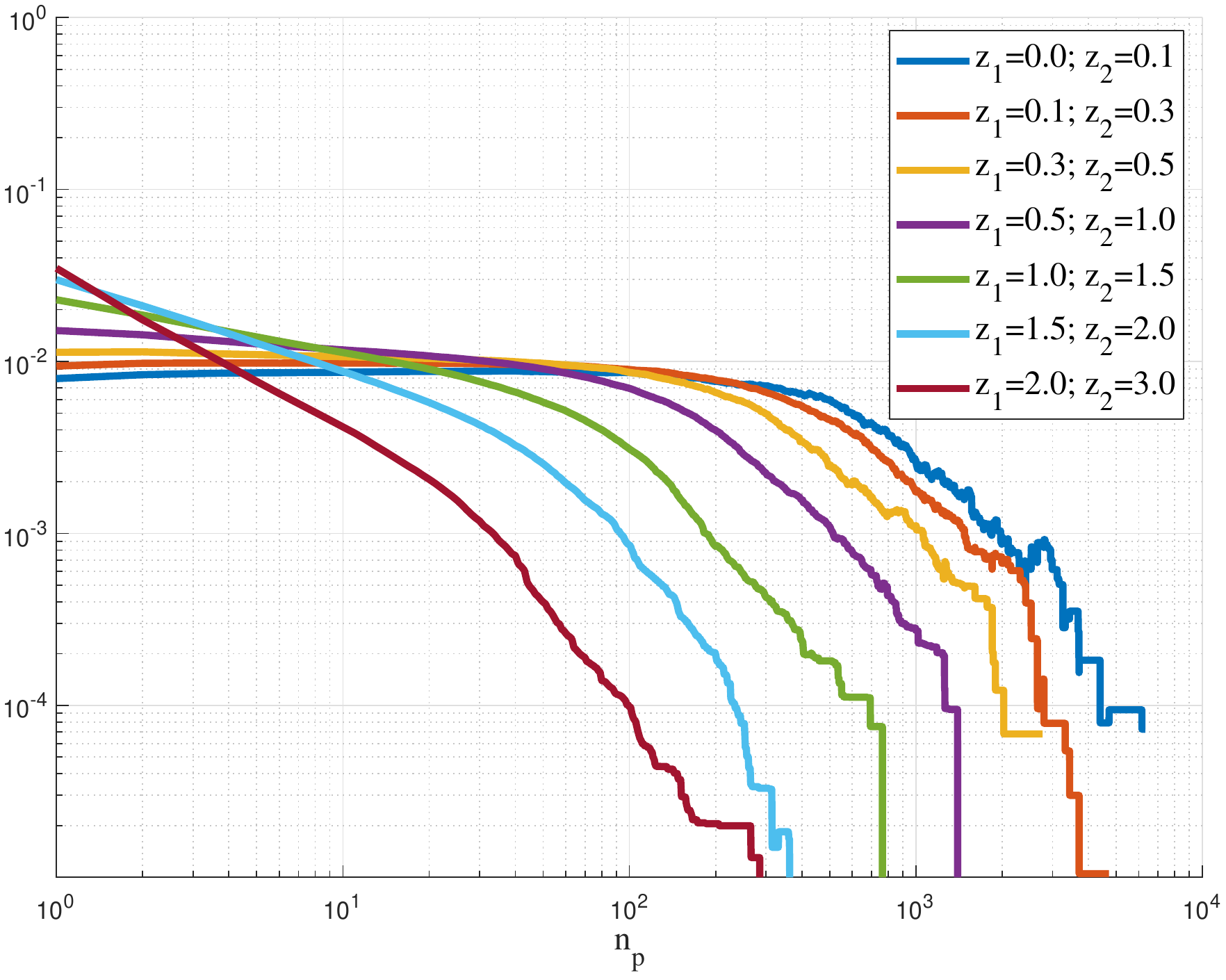}
\caption{The variation of flux function $\Pi _{ua} $ for halo rotational kinetic energy $K_{a}^{} \left(m_{h} ,a\right)$ with halo group size $n_{p} $. The flux function $\Pi _{ua} <0$ (inverse cascade from large to small scales) and is normalized by ${Nm_{p} u_{0}^{2}/t_{0} } $. A scale-independent flux function can be identified for mass propagation range with $m_{h} <m_{h}^{*} $.}
\label{fig:18}
\end{figure}

Figures \ref{fig:17} and \ref{fig:18} illustrate the variation of the flux functions for radial ($\Pi _{urp} $) and rotational kinetic energies ($\Pi _{ua} $) at different redshifts \citep[see datasets]{Xu:2022-Dark_matter-flow-dataset-part1,Xu:2022-Dark_matter-flow-dataset-part2}. Again, an inverse cascade is identified for both kinetic energies. The flux functions of radial and rotational motion are comparable to each other but is much smaller compared to the flux functions of halo virial and halo kinetic energies. From Eq. \eqref{ZEqnNum934449}, the flux function of any quantity is proportional to that quantity $V_{s}^{L} $ in typical halos. The rotational kinetic energy is estimated to be about 1\% of the specific potential in the same halo \citep[see][Table 3]{Xu:2022-The-mean-flow--velocity-disper}. The flux function of rotational kinetic energy is also about 1\% of the flux function of intra-halo potential, i.e. $\Pi _{ua} \approx 0.01\Pi _{\phi v} $ (see Fig. \ref{fig:8}). 

\section{Effect of halo shape on energy cascade}
\label{sec:5}

\subsection{The shape of halos and effect of shape change}
\label{sec:5.1}
The shape change of vortex plays a significant role for energy cascade in hydrodynamic turbulence. The shear-induced vortex stretching/elongating along the axis of rotation facilitates a direct energy cascade from large to small length scales (Eqs. \eqref{ZEqnNum426779} and \eqref{eq:2}). It should be interest to check if the shape change of halos, the counterpart of vortex, plays a similar role for energy cascade in SG-CFD. 

Halos are usually assumed to be spherical. The standard method to model halos as ellipsoids is to construct a 3x3 inertia tensor $I_{ij} $ for each halo \citep{Springel:2004-The-shapes-of-simulated-dark-m},
\begin{equation} 
\label{eq:67} 
I_{ij} =\sum _{p=1}^{n_{p} }x_{p,i}^{'} x_{p,j}^{'}  ,          
\end{equation} 
where $x_{p,i}^{'} $ is the \textit{i}th Cartesian component of position vector $\boldsymbol{\mathrm{x}}_{p}^{'} $ from the center of halo mass. The eigenvectors of inertia tensor $I_{ij} $ correspond to the directions of ellipsoid principle axis, while the eigenvalues ($r_{\lambda 1} ,r_{\lambda 2} ,r_{\lambda 3} $) correspond to the length of semimajor axis. The inertial tensor can be constructed for every halo identified in the system, where eigenvectors and eigenvalues can be determined for every halo identified. 

\begin{figure}
\includegraphics*[width=\columnwidth]{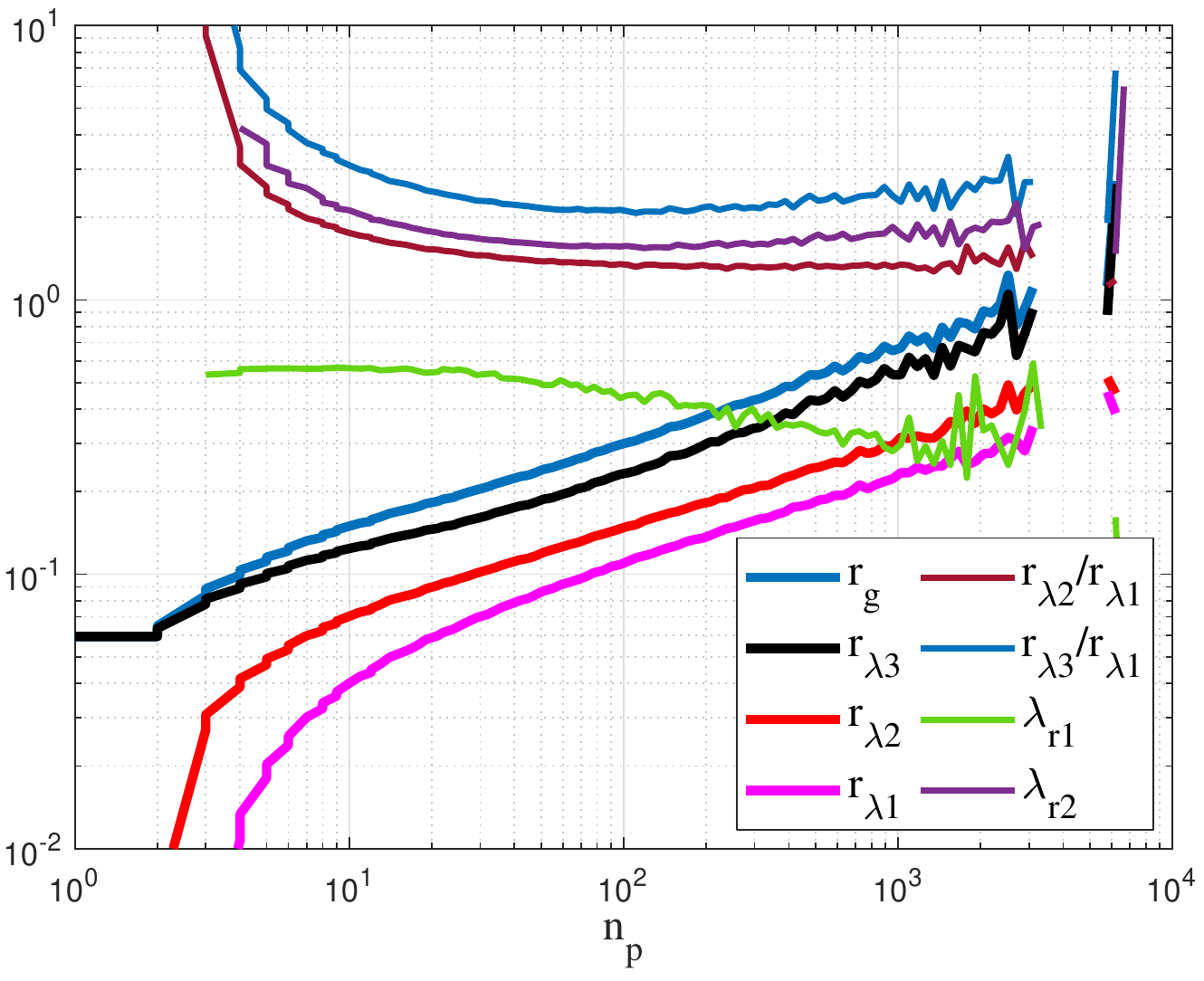}
\caption{The variation of mean square radius $r_{g} $ and three principle semiaxis of best fitting ellipsoid with $r_{\lambda 3} >r_{\lambda 2} >r_{\lambda 1}$ (Mpc/h) at \textit{z}=0. The ratio ${r_{\lambda 2}/r_{\lambda 1} } $rapidly decreases with halo size $n_{p} $ to a constant value of 1.35 at around a characteristic mass scale $n_{p}^{*} \approx 80$, while the ratio ${r_{\lambda 3}/r_{\lambda 1} } $decreases to a minimum values of 2 and the slowly increase to 2.8 for halos $n_{p} >n_{p}^{*} $. The value of $r_{\lambda 3} $ approaches $r_{g} $ for small halos. The first ratio (Eq. \eqref{ZEqnNum437512})$\lambda _{r1} \approx 0.5$ for small halos and $\lambda _{r1} \approx 0.3$ for large halos. The second ratio $\lambda _{r2} $ (Eq. \eqref{ZEqnNum262062}) is between [1.55 2], i.e the ellipsoid halos have a momentum of inertia about 1.55 to 2 times of that of a spherical halo with the same volume.}
\label{fig:19}
\end{figure}

Figure \ref{fig:19} presents the variation of length of three semimajor axis ($r_{\lambda 3} >r_{\lambda 2} >r_{\lambda 1}$) with halo group size $n_{p} $. The average is taken over all halos from the same group. The mean square radius $r_{g} $ is also presented in the same plot for comparison with the identity $r_{g}^{2} =r_{\lambda 1}^{2} +r_{\lambda 2}^{2} +r_{\lambda 3}^{2} $. The ratio ${r_{\lambda 2}/r_{\lambda 1} } $ rapidly decreases with halo size $n_{p} $ to a constant value of $\mathrm{\sim}$1.35 at around a characteristic mass scale $n_{p}^{*} \approx 80$, while the ratio ${r_{\lambda 3}/r_{\lambda 1} }$ decreases to a minimum value of 2 and then slowly rise to about 2.8 for halos $n_{p} >n_{p}^{*} $. 

The other two critical ratios $\lambda _{r1} $ and $\lambda _{r2} $ can be introduced and plotted in the same figure. The first ratio is defined as,
\begin{equation} 
\label{ZEqnNum437512} 
\lambda _{r1} =\frac{r_{\lambda 2} -r_{\lambda 1} }{r_{\lambda 3} -r_{\lambda 2} } ,           
\end{equation} 
where $\lambda _{r1} \approx 0.5$ is almost constant for small halos up to $n_{p} =40$ (see Eq. \eqref{ZEqnNum199101} and green line in Fig. \ref{fig:21} for explanation). This represents a unique evolution path (in mass space) of halo structure toward spherical shape. The second one is defined as the ratio of the moment of inertia of an ellipsoid rotating about the shortest semi-axis $r_{\lambda 1} $ to that of a sphere with the same volume, 
\begin{equation} 
\label{ZEqnNum262062} 
\lambda _{r2} =\frac{r_{\lambda 3}^{2} +r_{\lambda 2}^{2} }{2\left(r_{\lambda 1} r_{\lambda 2} r_{\lambda 3} \right)^{{2/3} } } ,          
\end{equation} 
where $\lambda _{r2} =1$ for spherical halos. Ellipsoid halo has a moment of inertia about $\lambda _{r2} =$1.55 to 2 times of that of a spherical halo with the same volume. The long-range gravity and tidal effects in SG-CFD might not be strong enough to deform halos significantly. Unlike the vortex stretching that can significantly change the momentum of inertia of vortex, the change of halo shape does not change the halo moment of inertia significantly and should not play a significant role in energy cascade. 

Note that vortex in incompressible flow is volume conserved with a uniform density. Halos are growing in mass and volume with a nonuniform density profile in an environment lack of incompressibility. These unique features lead to different dominant mechanisms that halo and vortex should play in energy cascade. In dark matter flow, halos cascade energy mostly facilitated by mass cascade (the growth of halo mass in Eq. \eqref{ZEqnNum934449}).

The shape of an ellipsoid can be systematically described by their triaxiality (prolate, oblate or triaxial), where a triaxiality parameter $h_{t} $ is introduced \citep{Franx:1991-The-Ordered-Nature-of-Elliptic}
\begin{equation} 
\label{eq:70} 
h_{t} =\frac{r_{\lambda 3}^{2} -r_{\lambda 2}^{2} }{r_{\lambda 3}^{2} -r_{\lambda 1}^{2} } ,           
\end{equation} 
which quantifies whether a halo is prolate ($h_{t} =1$) or oblate ($h_{t} =0$). 

Other important halo shape parameters that can be derived from three principle eigenvalues are the ellipticity $h_{e} $ and the prolateness $h_{p} $ \citep{Bardeen:1986-The-Statistics-of-Peaks-of-Gau,Despali:2014-Some-like-it-triaxial--the-uni}
\begin{equation}
h_{e} =\frac{r_{\lambda 3}^{} -r_{\lambda 1}^{} }{2\left(r_{\lambda 1}^{} +r_{\lambda 2}^{} +r_{\lambda 3}^{} \right)} \quad \textrm{and} \quad h_{p} =\frac{r_{\lambda 3}^{} -2r_{\lambda 2}^{} +r_{\lambda 1}^{} }{2\left(r_{\lambda 1}^{} +r_{\lambda 2}^{} +r_{\lambda 3}^{} \right)},   \label{eq:71}
\end{equation}

\noindent where $h_{p} =-h_{e} $ for oblate ellipsoid, $h_{p} =h_{e} $ for prolate ellipsoid. 

\begin{figure}
\includegraphics*[width=\columnwidth]{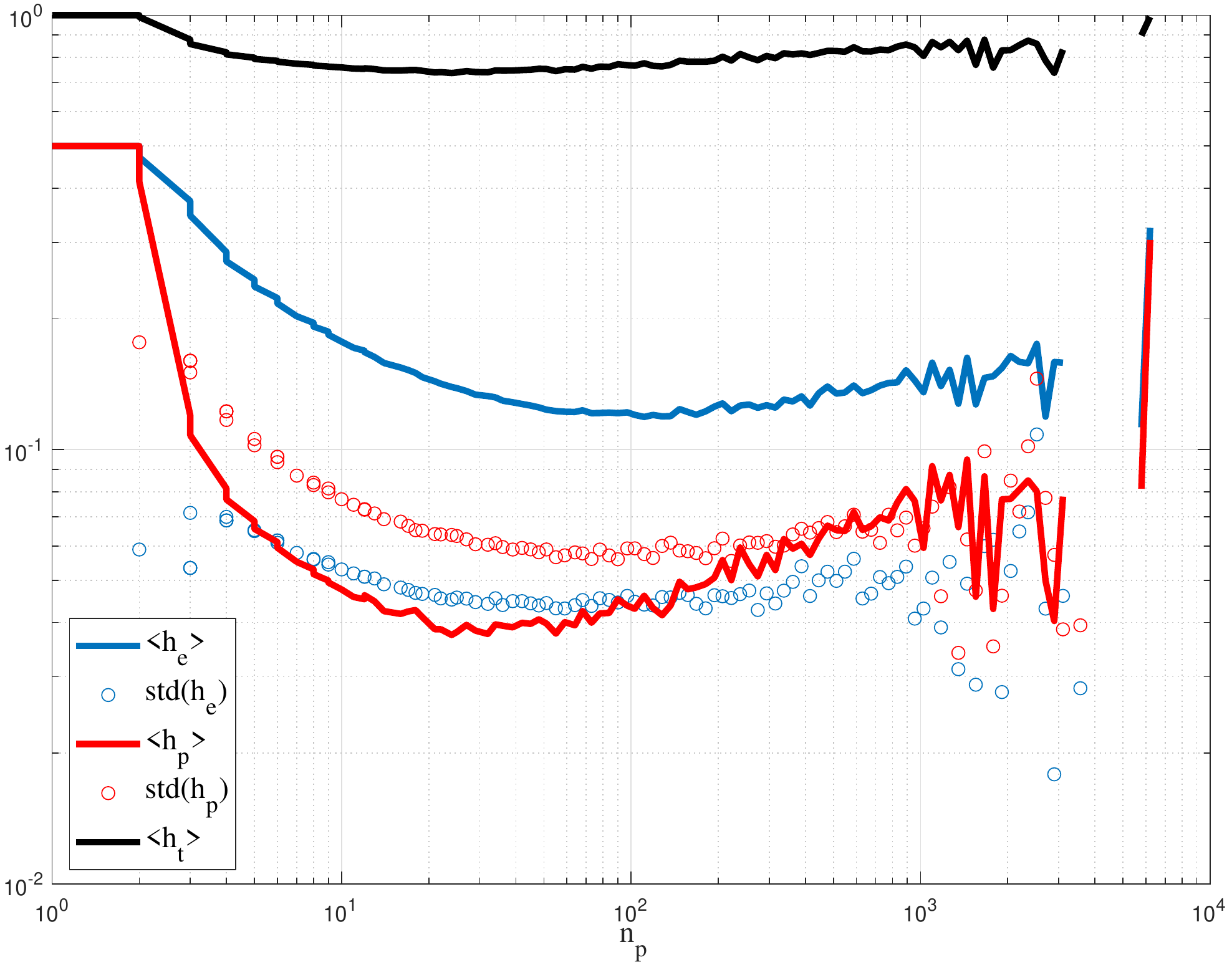}
\caption{The variation of mean and standard deviation of halo shape parameters (the ellipticity $h_{e} $ and prolateness $h_{p} $) with halo group size $n_{p} $ at z=0. Both shape parameters $\left\langle h_{e} \right\rangle $ and $\left\langle h_{p} \right\rangle $ have an initial sharp decrease from 0.5 for $n_{p} =2$ to a minimum value at around the characteristic halo size $n_{p}^{*} \approx 80$, followed by a slow increase with halo size for halos greater than $n_{p}^{*} $ that are not completely virialized. The triaxiality parameter $\left\langle h_{t} \right\rangle $ decreases from 1 for $n_{p} =2$ to 0.75 and slowly rises to 0.8$\mathrm{\sim}$0.9 for large halos.}
\label{fig:20}
\end{figure}

Figure \ref{fig:20} presents the variation of mean and standard deviation of halo shape parameters (the triaxiality $h_{t} $, ellipticity $h_{e} $, and prolateness $h_{p} $) with halo group size $n_{p}$ at \textit{z}=0. Both shape parameters $\left\langle h_{e} \right\rangle $ and $\left\langle h_{p} \right\rangle $ have an initial sharp drop from 0.5 for $n_{p} =2$ to a minimum value, followed by a slow increase with halo size for large halos that do not have enough time to completely virialize to a spherical shape. The mean triaxiality parameter $\left\langle h_{t} \right\rangle $ decreases from 1 for $n_{p} =2$ to 0.75 and slowly rises to 0.8$\mathrm{\sim}$0.9. The value of $h_{t}$ close to 1 indicates a prolate shape dominant for large halos. 

\subsection{Evolution of halo shape with size and redshift}
\label{sec:5.2}
\begin{figure}
\includegraphics*[width=\columnwidth]{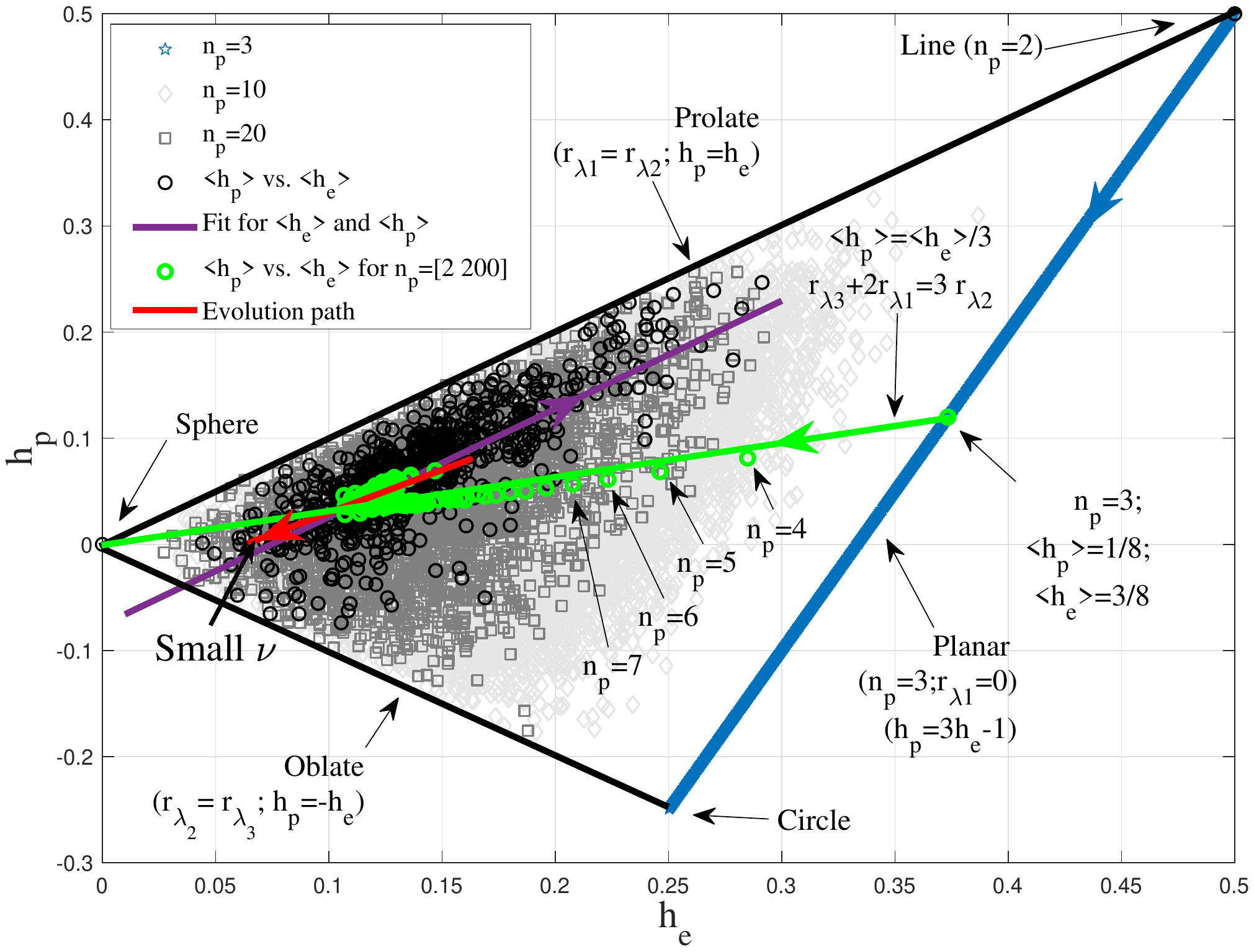}
\caption{The distribution of two shape parameter (the ellipticity $h_{e} $ and prolateness $h_{p} $) for different group size $n_{p} $ at \textit{z}=0. All $n_{p} =2$ halos collapse on to the top right corner with $h_{e} =h_{p} ={1/2} $. For all $n_{p} =3$ halos, the planar structure leads to a straight line $h_{p} =3h_{e} -1$ (blue line) with mean values of $h_{p} ={1/8} $ and $h_{e} ={3/8} $. The other two boundaries for $h_{e} $ and $h_{p} $ are also plotted as black lines for oblate and prolate ellipsoids, respectively. The mean shape parameters $\left\langle h_{e} \right\rangle $ and $\left\langle h_{p} \right\rangle $ for all halo groups of different sizes are presented as black circles (from Fig. \ref{fig:20}), while green circles highlight the halos in range of $n_{p} =\left[2,200\right]$. With increasing in size, the shape of halos evolves toward spherical structure along a unique path (green line: $\left\langle h_{p} \right\rangle ={\left\langle h_{e} \right\rangle/3} $) before a ``\textbf{V}'' turn. A fitted relation can be found $\left\langle h_{p} \right\rangle =\left\langle h_{e} \right\rangle -0.076$ (purple line) after ``\textbf{V}'' turn. The tip of ``\textbf{V}'' turn is identified as $\left\langle h_{e} \right\rangle =0.114$ and $\left\langle h_{p} \right\rangle =0.038$. Red line with arrow pointing to low peak height $\nu $ indicates the evolution path of halo shape from early to late stage.}
\label{fig:21}
\end{figure}

To better describe the evolution of halo shape for growing halos, we present $h_{e} $ and $h_{p} $ on the same plot. Figure \ref{fig:21} plots the distribution of two halo shape parameter (the ellipticity $h_{e} $ and prolateness $h_{p} $) for several different halo group sizes $n_{p} =2,3,10,20$ at \textit{z}=0. For the smallest halos $n_{p} =2$ with a line structure, all halos collapse onto the top right corner with $h_{e} =h_{p} ={1/2} $. For halos $n_{p} =3$ with a planar structure, all halos collapse onto a straight line $h_{p} =3h_{e} -1$ (blue line) with mean values of $h_{p} ={1/8} $ and $h_{e} ={3/8} $ for typical three-particle halos. The other two boundaries are also plotted as black lines for oblate ($h_{p} =-h_{e} $) and prolate ($h_{p} =h_{e} $) ellipsoids, respectively. With increasing halo size $n_{p} $, the distribution of $h_{e} $ and $h_{p} $ for larger halos ($n_{p} =10$ and $n_{p} =20$ in figure) shrinks and quickly approaches the prolate boundary. Obviously, halos exhibit a range of shapes with a preference for prolateness over oblateness \citep{Allgood:2006-The-shape-of-dark-matter-haloe,Tormen:1997-The-rise-and-fall-of-satellite}. This is expected as halos tend to form by collapsing along filaments and leading to prolateness. 

The mean shape parameters $\left\langle h_{e} \right\rangle $ and $\left\langle h_{p} \right\rangle $ for all halo groups in Fig. \ref{fig:20} are presented as black circles. To better describe how the halo structure evolves with halo size, we plot $\left\langle h_{e} \right\rangle $ and $\left\langle h_{p} \right\rangle $ for halo sizes $n_{p} $ between [2 200] (green circles). Clearly, the structure of halos evolves from $n_{p} =2$ to $n_{p} =3$ following the blue line (see arrows), and from $n_{p} =3$ to large halos around $n_{p}^{*} =80$ following the green line toward a spherical structure ($h_{p} =h_{e} =0$) with the equation,
\begin{equation}
\left\langle h_{p} \right\rangle ={\left\langle h_{e} \right\rangle/3} \quad  \textrm{or} \quad r_{\lambda 3} +2r_{\lambda 1} =3r_{\lambda 2},
\label{ZEqnNum199101}
\end{equation}
\noindent which leads to $\lambda _{r1} =0.5$ (Eq. \eqref{ZEqnNum437512} and Fig. \ref{fig:19}). 

This nearly perfect straight line of green circles clearly indicates: there exists a unique path for halo structure evolving from small size halos to characteristic size $n_{p}^{*} $ or $m_{p}^{*}$ that consistently approaches the spherical shape. For halos greater than the characteristic size $n_{p}^{*}$ with faster mass accretion, there may not be enough time to relax to equilibrium. Hence, the halo structure evolution takes a "\textbf{V}" turn and follows the purple line. A fitted relation $\left\langle h_{p} \right\rangle =\left\langle h_{e} \right\rangle -0.076$ (parallel to the prolate boundary) can be found. 

Evolution path of a given halo from early stage of its life with fast mass accretion (higher $\nu $) to late stage with a stable core and slow mass accretion (lower $\nu $) is presented in a separate paper \citep{Xu:2022-The-mean-flow--velocity-disper}, where $\nu ={\delta _{cr} /\sigma \left(m_{h} ,z\right)}$ is a reduced amplitude parameter (peak height) for density fluctuation. Here $\delta _{cr} \approx 1.68$ is the critical overdensity from spherical collapse model and $\sigma \left(m_{h} ,z\right)$ is the root-mean-square fluctuation of the smoothed density field. Specifically, the evolution of halo shape from early to late stage can be described by two shape parameters $h_{e} $ and $h_{p}$, both of which are functions of $\nu$ from N-body simulations \citep{Despali:2014-Some-like-it-triaxial--the-uni}, 
\begin{equation}
h_{e} =0.098\log _{10} \nu +0.094\quad \textrm{and} \quad h_{p} =0.079\log _{10} \nu +0.025,   
\label{ZEqnNum783026}
\end{equation}

The evolution path for a range of $\nu $ between [0.5 5] is also presented in the same plot (red line with arrow pointing to lower $\nu $). Halos at their early stage with fast mass accretion are primarily prolate with increasing angular momentum with time due to continuous mass accretion. The mass accretion and increase of angular momentum will gradually slower down when halos evolve to the late stage of their life along a path parallel to the prolate boundary (red line in Fig. \ref{fig:21}). Halos tend to be more spherical at their late stage. 

However, the change of moment of inertia from the change of halo shape should not be significant. The change of moment of inertia is mostly from mass accretion during that evolution \citep{Xu:2022-The-mean-flow--velocity-disper}. By contrast, the shear-induced vortex stretching in hydrodynamic turbulence involves a path along the prolate boundary from "sphere" to "line" (Fig. \ref{fig:21}), where moment of inertia changes significantly due to the change in shape.

\begin{figure}
\includegraphics*[width=\columnwidth]{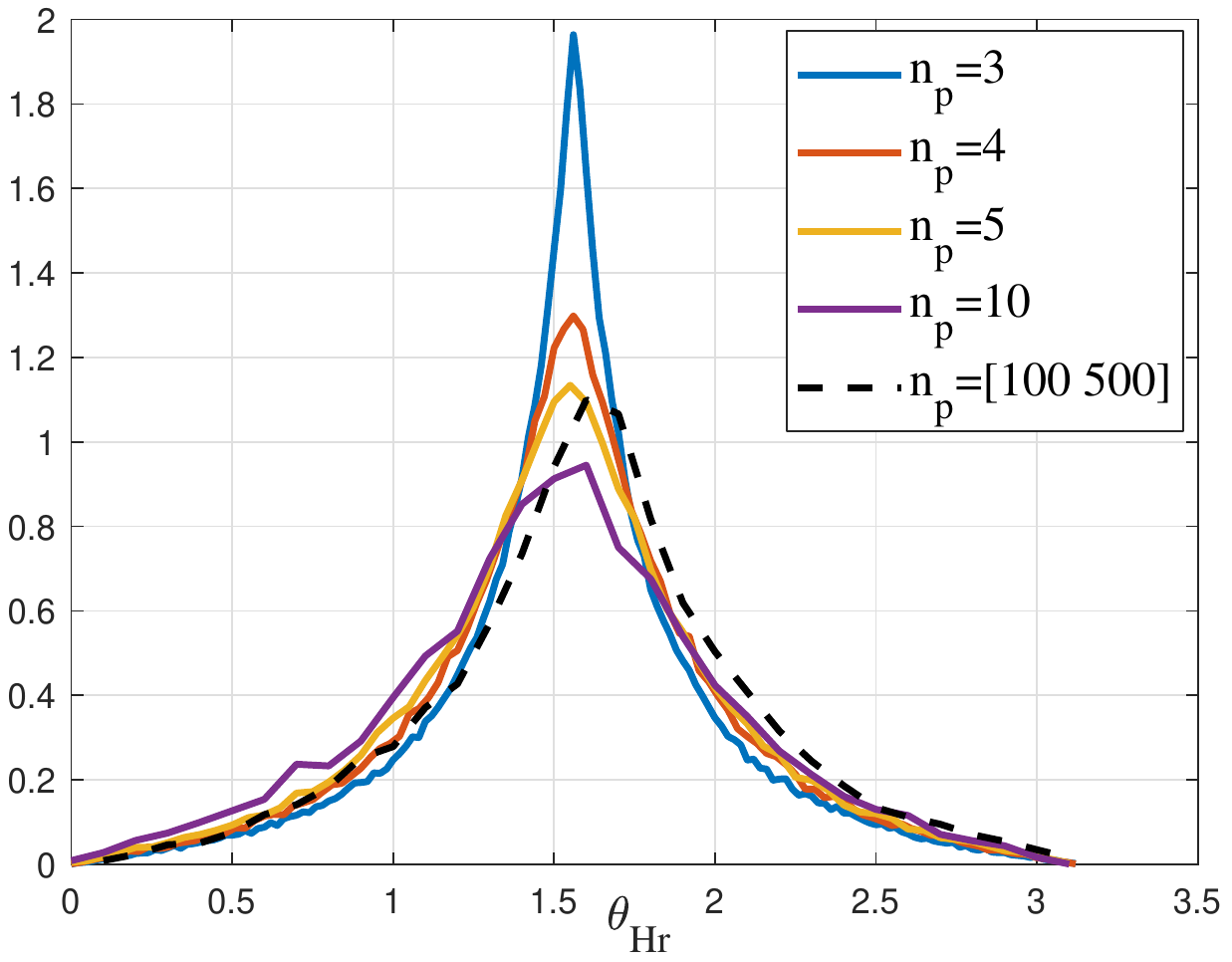}
\caption{The probability density function of angle $\theta _{Hr} $ between axis of rotation ($\boldsymbol{\mathrm{H}}_{h} $) and the major principal axis ($\boldsymbol{\mathrm{r}}_{\lambda 3} $) of triaxial ellipsoid halos. The axis of rotation of most halos is perpendicular to the major axis. By contrast, the axis of rotation of vortex is usually aligned with the major axis.} 
\label{fig:22}
\end{figure}

Finally, to explore the orientation of halo relative to the axis of rotation, Figure \ref{fig:22} plots the probability distribution of the angle $\theta _{Hr}$ between axis of rotation (direction of angular momentum $\boldsymbol{\mathrm{H}}_{h}$) and the major principle axis ($r_{\lambda 3}$) of triaxial ellipsoid halos for four halo groups of different sizes. The sharp peak around $\theta _{Hr} ={\pi/2}$ indicates that most halos have their spin axis lying perpendicular to their major axis. A wider and broader distribution of angle $\theta _{Hr} $ exists for larger halos. This finding agrees with previous results \citep{Allgood:2006-The-shape-of-dark-matter-haloe,Shaw:2006-Statistics-of-physical-propert,Bailin:2005-Internal-and-external-alignmen}. By contrast, the axis of rotation of vortex is usually aligned with the major axis. 

Halo interactions are local in mass space \citep{Xu:2021-Inverse-mass-cascade-mass-function}. The halo structure evolution is predominantly a continuous process in halo mass space via a sequential of merging with single mergers. It is reasonable to expect larger halos to have a strong memory of their progenitors that they inherit their mass from (the green path in Fig. \ref{fig:21}). For halos of $n_{p} =2$, the axis of rotation must be perpendicular to the major axis and this feature is largely inherited with increasing halo sizes (see Fig. \ref{fig:22} for halos with $n_{p} =3$, 4, 5 and 10). This also explains why halo tends to rotate around an axis perpendicular to its the major principal axis.

In hydrodynamic turbulence, the vortex is stretching along the axis of rotation that facilitates the direct energy cascade, while the volume of vortex is conserved due to incompressibility, i.e. vortex more likely evolves along the prolate boundary toward a line structure in Fig. \ref{fig:21}. In dark matter flow (SG-CFD), halos smaller than the characteristic size $m_{h}^{*} $ evolves along a unique path approaching the spherical shape (green line in Fig. \ref{fig:21}). For halos greater than $m_{h}^{*} $, the halo shape evolves parallel to the prolate boundary. The tip of "\textbf{V}" turn with $\left\langle h_{e} \right\rangle =0.114$ and $\left\langle h_{p} \right\rangle =0.038$ in Fig. \ref{fig:21}. The change of halo shape should not be the dominant mechanism for energy cascade. Due to the inverse mass cascade, both mass and volume of halos increases with time. Halo density is also nonuniform. The shape change of halos does not play a dominant role on the energy cascade, while the mass cascade should be the dominant mechanism that enables the energy cascade in SG-CFD (discussed in Section \ref{sec:4}). 

\section{Conclusions}
\label{sec:6}
By revisiting some fundamental ideas of energy cascade in hydrodynamic turbulence, the energy cascade in self-gravitating collisionless dark matter flow shares many similarities, but also exhibits many unique features. This paper focus on the energy cascade in dark matter flow and its dominant mechanism. 

The starting point is a halo-based description of collisionless system that can be divided into halo and out-of-halo sub-systems. The particle motion can be decomposed into the random motion of halos and the motion in halos. While virial equilibrium should be established for both types of motion, the equilibrium for motion in halos is established much earlier than the random motion of halos (Fig. \ref{fig:9}). Total energy of out-of-halo sub-system is conserved with time, where potential cancels kinetic energy with an effective virial ratio of 2 (Figs. \ref{fig:6} and \ref{fig:9}). The energy change in the entire system mostly comes from the virilization in halo sub-system. The total kinetic/potential energies in halo-sub system increase $\mathrm{\propto}$\textit{t} with a (mass) scale- and time-independent flux function ($\epsilon_u$) in mass propagation range once the statistically steady state is established (Figs. \ref{fig:10} and \ref{fig:11}). A continuous mass exchange between two sub-systems at the smallest scale is required to sustain the growth of total halo mass as $M_h\propto$a$^{1/2}$ and the linear growth of total energy in all halos $E\propto$a$^{3/2}$ (Figs. \ref{fig:9}-\ref{fig:11}). 

The mass exchange between two sub-systems drives the inverse mass cascade in halo sub-system. By introducing the energy flux and transfer functions, an inverse cascade of kinetic energy can be identified with kinetic energy transferred from small to large mass scales, and vice versa for potential energy with a direct cascade (Figs. \ref{fig:3}, \ref{fig:4}, \ref{fig:7}, and \ref{fig:8}). Both cascades are related by the virial theorem. The direct cascade of potential energy to the smallest scale sustains the inverse cascade of kinetic energy. Both potential and kinetic energies exhibit a scale- and time-independent flux in the mass propagation range with energy flux proportional to mass flux (Eqs. \eqref{ZEqnNum894335} and \eqref{ZEqnNum316450}). 

The coherent motion (mean flow) in halos includes the radial and rotational motion. By correctly modeling the dependence of halo radial and angular momentum, and angular velocity on halo mass and redshift (Eqs. \eqref{ZEqnNum671144}, \eqref{ZEqnNum316976}, and \eqref{ZEqnNum528733} and \eqref{ZEqnNum523492} and Figs. \ref{fig:12}, \ref{fig:14} , \ref{fig:15} and \ref{fig:16}), an inverse cascade can also be identified for kinetic energy due to coherent radial and rotational motion in halos (Figs. \ref{fig:17} and \ref{fig:18}). For a given halo mass, the rotational kinetic energy is independent of time, while the radial (peculiar) kinetic energy decreases with time (Eqs.    \eqref{ZEqnNum528733} and \eqref{ZEqnNum523492}). 

The flux function of a general specific quantity is also formulated (Eqs. \eqref{ZEqnNum593887} and \eqref{ZEqnNum802989}). In SG-CFD, the energy cascade can be quantitatively described by the mass accretion of typical halos from small to large mass scales (Eq. \eqref{ZEqnNum934449}). By contrast, the vortex stretching/elongating along its axis of rotation facilitates a direct energy cascade from large to small length scales in hydrodynamic turbulence. The vortex is volume conserved with a uniform density due to the incompressibility. The axis of rotation is usually aligned with the major axis of vortex.

The shape evolution of halos is also extensively studied for SG-CFD. The halo structure evolution is continuous in mass space with halos inheriting structures from their progenitors. In mass space, the shape of halos evolves along a unique path gradually approaching the spherical shape with increasing halo size (green line in Fig. \ref{fig:21}). Large halos exhibit a range of shapes with a preference for prolateness over oblateness. Most halos have their spin axis perpendicular to their major axis (Fig. \ref{fig:22}). In redshift space, halos are more prolate in their early stage of life (fast mass accretion) and become more spherical in their late stage (slow mass accretion) (see red line in Fig. \ref{fig:21} and Eq. \eqref{ZEqnNum783026}). However, change of halo shape should not play a significant role in energy cascade. Change in halo moment of inertial due to the change in shape is small and less than 2 times (Eq. \eqref{ZEqnNum262062} and Fig. \ref{fig:19}). 

Finally, this paper focus on the energy transfer and cascade across halos of different mass scales, while future study should extend to the energy transfer within individual halos that requires complete solutions of the mean flow and velocity dispersion \citep{Xu:2022-The-mean-flow--velocity-disper}.


\section*{Data Availability}
Two datasets underlying this article, i.e. a halo-based and correlation-based statistics of dark matter flow, are available on Zenodo \citep{Xu:2022-Dark_matter-flow-dataset-part1,Xu:2022-Dark_matter-flow-dataset-part2}, along with the accompanying presentation slides "A comparative study of dark matter flow \& hydrodynamic turbulence and its applications" \citep{Xu:2022-Dark_matter-flow-and-hydrodynamic-turbulence-presentation}. All data files are also available on GitHub \citep{Xu:Dark_matter_flow_dataset_2022_all_files}.

\bibliographystyle{mnras}
\bibliography{Papers}

\appendix

\label{lastpage}
\end{document}